\documentclass[preprintnumbers, prd, showpacs, floatfix,onecolumn, superscriptaddress,nofootinbib]{revtex4}
\usepackage{amsfonts}  %Revised  in 27 Aug 2017%
\usepackage{amsmath}
\usepackage{amssymb,epsf}
\usepackage{latexsym}
\usepackage{graphicx,epsfig,color}
\usepackage{amssymb}
\usepackage{subfigure}
\usepackage{epstopdf}
\usepackage{float}
\begin{document}
\title{Critical behavior and phase transition of dilaton black holes
with nonlinear electrodynamics}

\author{Z. Dayyani$^{1}$,  A. Sheykhi $^{1,2}$\footnote{
asheykhi@shirazu.ac.ir}, M. H. Dehghani$^{1}$\footnote{
mhd@shirazu.ac.ir} and S. Hajkhalili}
\address{$^1$ Physics Department and Biruni Observatory, College of
Sciences, Shiraz University, Shiraz 71454, Iran\\
$^2$Research Institute for Astronomy and Astrophysics of Maragha
(RIAAM), P.O. Box 55134-441, Maragha, Iran}

\begin{abstract}
In this paper, we take into account the dilaton black hole
solutions of Einstein gravity in the presence of logarithmic and
exponential forms of nonlinear electrodynamics. At first, we
consider the cosmological constant and nonlinear parameter as
thermodynamic quantities which can vary. We obtain thermodynamic
quantities of the system such as pressure, temperature and Gibbs
free energy in an extended phase space. We complete the analogy of
the nonlinear dilaton black holes with Van der Waals liquid-gas
system. We work in the canonical ensemble and hence we treat the
charge of the black hole as an external fixed parameter. Moreover,
we calculate the critical values of temperature, volume and
pressure and show they depend on dilaton coupling  constant as
well as nonlinear parameter. We also investigate the critical
exponents and find that they are universal and independent of the
dilaton and nonlinear parameters, which is an expected result.
{Finally, we explore the phase transition of nonlinear dilaton
black holes by studying the Gibbs free energy of the system. We
find that in case of $T>T_c$, we have no phase transition. When
$T=T_c$, the system admits a second order phase transition, while
for $T=T_{\rm f}<T_c$ the system experiences a first order
transition. Interestingly, for $T_{\rm f}<T<T_c$ we observe a
\textit{zeroth order} phase transition in the presence of dilaton
field. This novel \textit{zeroth order} phase transition is
occurred due to a finite jump in Gibbs free energy which is
generated by dilaton-electromagnetic coupling constant, $\alpha$,
for a certain range of pressure. }


\end{abstract}
 \maketitle

\section{Introduction}
Nowadays , it is a general belief that there should be some deep
connection between gravity and thermodynamics. Bekenstein
\cite{Bek} was the first who disclosed that black hole can be
regarded as a thermodynamic system with entropy and temperature
proportional, respectively, to the horizon area and surface
gravity \cite{Bek,Haw}. The  temperature   $T$  and entropy $S$
together with the energy (mass) of the black holes satisfy the
first law of thermodynamics $dM=TdS$ \cite{Bek,Haw}. Historically,
Hawking and Page were the first who reported the existence of a
certain phase transition in the phase space of the Schwarzschild
anti-de Sitter (AdS) black hole \cite{hawking-page}. In recent
years, the studies on the phase transition of gravitational
systems have got a renew interest. It has been shown that one can
extend the thermodynamic phase space of a Reissner-Nordstrom (RN)
black holes in an AdS space, by
considering the cosmological constant as a thermodynamic pressure, $%
P=-\Lambda /8\pi $ and its conjugate quantity as a thermodynamic volume \cite%
{Do1,Ka,Do2,Do3,Ce1,Ur}. In particular, it was argued that indeed
there is a complete analogy for RN-AdS black holes with the van
der Walls liquid-gas system with the same critical exponents
\cite{MannRN}. The studies were also extended to nonlinear
Born-Infeld electrodynamics \cite{MannBI}. In this case, one needs
to introduce a new thermodynamic quantity conjugate to the
Born-Infeld parameter which is required for consistency of both
the first law of thermodynamics and the corresponding Smarr
relation \cite{MannBI}. Extended phase space thermodynamics and
P-V criticality of black holes with power-Maxwell electrodynamics
were investigated in \cite{HV}. When the gauge field is in the
form of  logarithmic and exponential nonlinear electrodynamics,
critical behaviour of  black hole solutions in Einstein gravity
have also been explored \cite{Hendi1}. Treating the cosmological
constant as a thermodynamic pressure, the effects of higher
curvature corrections from Lovelock gravity on the phase structure
of asymptotically AdS black holes have also been explored. In this
regards, critical behaviour and phase transition of higher
curvature corrections such as Gauss-Bonnet \cite{GB1,GB2} and
Lovelock gravity have also been investigated \cite{Lovelock}. The
studies were also extended to the rotating black holes, where
phase transition and critical behavior of Myers-Perry black holes
have been investigated \cite{Sherkat}. Other studies on the
critical behavior of black hole spacetimes in an extended phase
space have been carried out in \cite{Sherkat1,Rabin,Zou,John}.

Although Maxwell theory is able to explain varietal phenomena in
electrodynamics, it suffers some important problems such as
divergency of the electric field of a point-like charged particle
or infinity of its self energy. In order to solve these problems,
one may get help from the nonlinear electrodynamics
\cite{Born,Soleng, Hassaine,Hendi3}. Inspired by developments in
string/M-theory, the investigation on the nonlinear
electrodynamics has got a lot of attentions in recent years.

On the other side, a scalar field called dilaton emerges in the
low energy limit of string theory \cite{Green}. Breaking of
space-time supersymmetry in ten dimensions, leads to one or more
Liouville-type potentials, which exist in the action of dilaton
gravity. In addition, the presence of the dilaton field is
necessary if one couples the gravity to other gauge fields.
Therefore,  the dilaton field plays an essential role in string
theory and it has  attracted extensive attention in the
literatures \cite{d1,d2,CHM,d4,d5,Cai3,neda,Shey3,d7}. Critical
behavior of the Einstein-Maxwell-dilaton black holes has been
studied in \cite{Kamrani}. In the context of Born-Infeld and
power-Maxwell nonlinear electrodynamics coupled to the dilaton
field, critical behavior of $(n+1)$-dimensional topological black
holes in an extended phase space have been explored in
\cite{Dayyani1} and \cite{Dayyani2}, respectively. Although, the
asymptotic behavior of these solutions \cite{Dayyani1,Dayyani2}
are neither flat nor ant-de Sitter (AdS), it was found that the
critical exponents have the universal mean field values and do not
depend on the details of the system, while thermodynamic
quantities depend on the dilaton coupling constant, nonlinear
parameter and the dimension of the spacetime. In the present work,
we would like to extend the study on the critical behaviour of
black holes, in an extended phase space, to other nonlinear
electrodynamics in the context of dilaton gravity such as
exponential and logarithmic nonlinear electrodynamics. Following
\cite{MannBI,Dayyani2}, and in order to satisfy the Smarr
relation, we shall extend the phase space to include nonlinear
parameter as a thermodynamic variable and consider it`s conjugate
quantity as polarization. We will complete analogy of the
nonlinear dilaton black holes with Van der Waals liquid-gas system
and work in the canonical ensemble. In addition, we calculate the
critical exponents and show that they are universal and are
independent of the dilaton and nonlinearity parameters. {Finally,
we shall explore the phase transition of dilaton black holes
coupled to nonlinear electrodynamics by considering the
discontinuity in the Gibss free energy of the system. We will see
that in addition to the first and second order phase transition in
charged black holes, the presence of the dilaton field admits a
\textit{zeroth order} phase transition in the system. This phase
transition is occurred due to a finite jump in Gibbs free energy
which is generated by dilaton-electromagnetic coupling constant,
$\alpha$, for a certain range of pressure. This novel behavior
indicates a small/large black hole \emph{zeroth-order} phase
transition in which the response functions of black holes
thermodynamics diverge e.g. isothermal compressibility.}

This paper is outlined as follows. In the next section, we present
the action, basic field equations and our metric ansatz for
dilaton black holes. In section \ref{END}, we explore the
critical behaviour of dilaton black holes coupled to exponential
nonlinear (EN) electrodynamics. In section \ref{LND}, we
investigate $P-V$ criticality of dilaton black holes when the
gauge field is in the form of logarithmic nonlinear (LN)
electrodynamics. In section \ref{Effect}, we investigate the
effects of nonlinear gauge field parameter in the strong nonlinear
regime on the critical behaviour of the system. {In section
\ref{PT}, we explore the phase transition of nonlinear dilaton
black holes.} We finish with closing remarks in section
\ref{Clos}.

\section{Basic field equations}\label{Field}
We examine the following action of Einstein-dilaton gravity which
is coupled to nonlinear electrodynamics,
\begin{equation}\label{Act}
S=\frac{1}{16\pi}\int{d^{4}x\sqrt{-g}\left(\mathcal{R}\text{ }-2
g^{\mu\nu}
\partial_{\mu} \Phi \partial_{\nu}\Phi -V(\Phi
)+L(F,\Phi)\right)},
\end{equation}
where $\mathcal{R}$ is the Ricci scalar curvature, $\Phi $ is the
dilaton field and $V(\Phi )$ is the potential for $\Phi $. We
assume the dilaton potential in the form of two Liouville terms
\cite{CHM,Shey3}
\begin{equation}\label{pot}
V(\Phi) = 2\Lambda_{0} e^{2\zeta_{0}\Phi} +2 \Lambda e^{2\zeta
\Phi},
\end{equation}
where $\Lambda_{0}$,  $\Lambda$, $ \zeta_{0}$ and $ \zeta$ are
constants that should be determined. In action (\ref{Act}),
$L(F,\Phi)$ is the Lagrangian of two Born-Infeld likes nonlinear
electrodynamics which are coupled to the dilaton field
\cite{somayeh,sara}
\begin{equation}\label{lag}
L(F,\Phi)=\left\{
  \begin{array}{ll}
    $$4\beta^{2} e^{2\alpha \Phi}\left[\exp\left(-\frac{e^{-
    4\alpha \Phi}F^2}{4\beta^{2}}\right)-1\right],\quad \quad\quad $$\rm END , &  \\
    &\\
$$ -8\beta^2 e^{2\alpha\Phi} \ln \left(1+\frac{ e^{-4\alpha\Phi}F^2}
{8 \beta^2}\right),\quad~ \quad\quad \quad$$ \rm LND,
  \end{array}
\right.
\end{equation}
where END and LND stand for exponential and logarithmic nonlinear
dilaton Lagrangian, respectively. Here $\alpha$ is a constant
which determines the strength of coupling of dilaton and
electromagnetic field. The parameter $\beta$ with dimension of
mass, represents the maximal electromagnetic field strength which
in string theory can be related to the \textit{string tension},
$\beta=\frac{1}{2\pi \alpha'}$\cite{GW}. In fact $\beta$
determines the strength of the nonlinearity of the
electrodynamics. In the limit of large $\beta$
($\beta\rightarrow\infty$), the systems goes to the linear regime
and the nonlinearity of the theory disappears and the nonlinear
electrodynamic theory reduces to the linear Maxwell
electrodynamics. On the other hand, as $\beta$ decreases
($\beta\rightarrow0$), we go to the strong nonlinear regime of the
electromagnetic and thus the behavior of the system will be
completely different (see section V of the paper). In expression
(\ref{lag}) $F^2=F_{\mu \nu }F^{\mu \nu }$, where $F_{\mu \nu }$
is the electromagnetic field tensor. By varying action (\ref{Act})
with respect to the gravitational field $g_{\mu \nu }$, the
dilaton field $\Phi $ and the electromagnetic field $A_{\mu }$, we
arrive at the following field equations \cite{somayeh,sara}
\begin{eqnarray}\label{FE1}
{\cal R}_{\mu\nu}&=& 2 \partial _{\mu }\Phi
\partial _{\nu }\Phi +\frac{1}{2}g_{\mu \nu }V(\Phi)-
2e^{-2\alpha \Phi}\partial_{Y}{{\cal L}}(Y) F_{\mu\eta}
F_{\nu}^{~\eta } \nonumber \\
&&+n\beta^2 e^{2\alpha \Phi} \left[2Y\partial_{Y}{{\cal
L}}(Y)-{{\cal L}}(Y)\right]g_{\mu\nu},
\end{eqnarray}
\begin{equation}\label{FE2}
\nabla ^{2}\Phi =\frac{1}{4}\frac{\partial V}{\partial \Phi}+
n\alpha \beta^2 e^{2\alpha \Phi }\left[2{ Y}\partial_{Y}{{\cal
L}}(Y)-{\cal L}(Y)\right],
\end{equation}
\begin{equation}\label{FE3}
\nabla _{\mu }\left(e^{-2\alpha \Phi}
\partial_{Y}{{\cal L}}(Y) F^{\mu\nu}\right)=0,
\end{equation}
where $n=2$ for END and $n=-4$ for LND cases. In the above field
equations we have used a shorthand for $L(F,\Phi)$ as
\begin{eqnarray}\label{lag2}
&&L(F,\Phi)=2 n\beta^2{\mathcal{L}}(Y),~~~~~~~~~{\mathcal{L}}(Y)=\left\{
  \begin{array}{ll}
    $$\exp(-Y)-1,\quad \quad\qquad $$\rm END  &  \\
    &\\
    $$ \ln(1+Y),\quad \quad\quad ~\qquad$$\rm LND
  \end{array}
\right.
\end{eqnarray}
and
\begin{eqnarray}\label{Y}
Y= \frac{e^{-4\alpha \Phi}F^2}{2 \beta^2|n|}.
\end{eqnarray}
In the limiting case $\beta\rightarrow\infty$, which is equal to
${\mathcal{L}}(Y)=-Y$ for END and ${\mathcal{L}}(Y)=Y$ for LND
cases, the above system of equations recover the corresponding
equations for Einstein-Maxwell-dilaton gravity \cite{Shey3}.

We would like to find topological solutions of the above field
equations. The most general such metric can be written in the form
\begin{equation}\label{metric}
ds^2=-f(r)dt^2 +{dr^2\over f(r)}+ r^2R^2(r)d\Omega_{k}^2 ,
\end{equation}
where $f(r)$ and $R(r)$ are functions of $r$ which should be
determined, and $d\Omega_{k}^2$ is the line element of a
two-dimensional hypersurface $\Sigma$ with constant curvature,
\begin{equation}\label{met}
d\Omega_k^2=\left\{
  \begin{array}{ll}
    $$d\theta^2+\sin^2\theta d\phi^2$$,\quad \quad\!\!{\rm for}\quad $$k=1$$, &  \\
    $$d\theta^2+\theta^2 d\phi^2$$,\quad\quad\quad {\rm for}\quad $$k=0$$,&  \\
    $$d\theta^2+\sinh^2\theta d\phi^2$$, \quad {\rm for}\quad $$k=-1$$.&
  \end{array}
\right.
\end{equation}
For $k = 1$, the topology of the event horizon is the two-sphere
$S^2$, and the spacetime has the topology $R^2 \times S^2$. For $k
= 0$, the topology of the event horizon is that of a torus and the
spacetime has the topology $R^2 \times T^2$. For $k = -1$, the
surface $\Sigma$ is a $2$-dimensional hypersurface $H^2$ with
constant negative curvature. In this case the topology of
spacetime is $R^2 \times H^2$.

In the remaining part of this paper, we consider the critical
behaviour of END  and LND black holes.
%%%%%%%%%%%%%%%%%%%%%%%%%%%%%%%%%%%%%%%%%%%%%%%%%%%%%%%%%%%%%%%%%%%%%%%%%%%%%%%%%%%%%%%%%%%
\section{Critical behavior of END black holes} \label{END}
In this section, at first, we review the solution of dilatonic
black holes coupled to EN electrodynamics \cite{somayeh}. Then, we
construct Smarr relation and equation of state of the system to
study the critical behavior of the system.
\subsection{Review on END black holes}
In order to solve the system of equations (\ref{FE1}) and
(\ref{FE2}) for three unknown functions $f(r)$, $R(r)$ and $\Phi
(r)$, we make the ansatz \cite{neda}
\begin{equation}
R(r)=e^{\alpha \Phi}.\label{Rphi}
\end{equation}
Inserting this ansatz and metric (\ref{metric}) into the field
equations (\ref{FE1})-(\ref{FE3}), one can show that these
equations have the following solutions \cite{somayeh}
\begin{equation}\label{phi}
\Phi (r)=\frac{\alpha }{\alpha ^{2}+1}\ln
\left(\frac{b}{r}\right),
\end{equation}
\begin{eqnarray}\label{vectorpot}
A_{t}^{\rm END}&=&b^{\gamma}\beta(\alpha^2+1)\left(\frac{\beta
b^{\gamma}}{q}\right)^{\frac{1-\gamma}{\gamma-2}}\left(\frac{1-\alpha^2}{4}\right)^{\frac{1}{2\gamma-4}}\nonumber
\\ && \times \Bigg{\{}
-\frac{1}{4}\Gamma\left(\frac{\alpha^2+1}{4},\frac{1-\alpha^2}{4}L_{W}(\eta)\right)
+\frac{1}{\alpha^2-1}\left[\Gamma\left(\frac{\alpha^2+5}{4},\frac{1-\alpha^2}{4}L_{W}(\eta)\right)
-\frac{1}{2}\Gamma\left(\frac{\alpha^2+1}{4}\right)\right]
\Bigg{\}}.\nonumber \\
\end{eqnarray}
\begin{eqnarray}\label{fEND}
f(r)\mid_{\rm END}&=&-k\frac{ {\alpha}^{2}+1}{ {\alpha}^{2}-1
}{b}^{-\gamma}{r}^{\gamma}-\frac{m}{r^{1-\gamma}}+\frac{(\Lambda+2\beta^2)
\left( {\alpha}^{2}+1 \right) ^{2}{b}^{\gamma}}{\alpha^{2}-3
}r^{2-\gamma}+2 \beta q{\left(\alpha^2+1\right)^2}
r^{\gamma-1}{\left(\frac{\beta^2
b^{2\gamma}}{q^2}\right)}^{\frac{1-\gamma}{2\gamma-4}}{L_W^{\frac{3-2\gamma}{2\gamma-4}}(\eta)}
\nonumber
\\ &&\times\Bigg{\{}\frac{L_W^2(\eta)}{\alpha^2+5}\digamma\left(\frac{\alpha^2+5}{4},
\frac{\alpha^2+9}{4},\frac{\alpha^2-1}{4}L_W(\eta)\right)
-
\frac{1}{\alpha^2-3}\digamma\left(\frac{\alpha^2-3}{4},\frac{\alpha^2+1}{4},\frac{\alpha^2-1}{4}L_W(\eta)\right)
\Bigg{\}}, \nonumber\\
\end{eqnarray}
where $m$ and $q$ are integration constants which are related to
the mass and the charge of the black holes. Also,
$L_W(x)={LambertW(x)}$ is Lambert function and $ \digamma(a,b,z)$
is the hypergeometric function \cite{Lambert}. Here $\eta$ and
$\gamma$ have definition as
\begin{equation}
\eta\equiv \frac{q^{2}r^{2\gamma-4}}{\beta ^{2}b^{2\gamma
}},\qquad\gamma=\frac{2\alpha^2}{\alpha^2+1}. \label{eta}
\end{equation}
The above solutions will fully satisfy the system of equations
(\ref{FE1}) and (\ref{FE2}) provided we have
\begin{eqnarray}
\zeta_{0}=\frac{1}{\alpha},\qquad
\zeta=\alpha,\qquad\Lambda_0=k\frac{b^{-2}\alpha^2}{\alpha^2-1}.
\end{eqnarray}
According to the definition of mass due to Abbott and Deser
\cite{abot}, the mass of the solution (\ref{fEND}) is
\cite{somayeh}
\begin{equation}
{M}=\frac{b^{\gamma}m \omega}{8\pi(\alpha^2+1)}.\label{Mass}
\end{equation}
where $\omega$ represents the area of the constant hypersurface
$\Sigma$. In relation (\ref{Mass}), one can find mass parameter as
a function of horizon radius by considering $f(r=r_+)=0$
\cite{somayeh}. The charge of the solution is given by
\cite{somayeh}
\begin{equation}
{Q}=\frac{q\omega}{4\pi},  \label{Charge}
\end{equation}
The Hawking temperature of END black hole can be calculated as
\cite{somayeh}
\begin{eqnarray}\label{TemEND}
T_{+}\mid_{END}=\frac{1}{4\pi}\left(\frac{df(r)}{dr}\right)_{r_+}=-\frac{(\alpha^2+1)}{4\pi}{r_{+}
^{1-\gamma}}\Bigg{\{}k\frac{ b^{-\gamma}
r_{+}^{2\gamma-2}}{\alpha^2-1}+(\Lambda+2\beta^2)b^{\gamma}-2\beta
q
r_{+}^{\gamma-2}\left(\frac{1}{\sqrt{L_{W}(\eta_{+})}}-\sqrt{L_{W}(\eta_{+})}\right)\Bigg{\}},
\nonumber\\
\end{eqnarray}
where $\eta_+=\eta({r=r_+})$. Applying the well-known area law, we
can find entropy of black hole as
\begin{equation}\label{entropy}
S=\frac{A}{4}=\frac{b^\gamma r_+^{2-\gamma}\omega}{4}.
\end{equation}
The electric potential of the black hole is obtained as
\cite{somayeh}
\begin{eqnarray}\label{potEND}
&&U\mid_{\rm END}=b^{\gamma}\beta(\alpha^2+1)\left(\frac{\beta
b^{\gamma}}{q}\right)^{\frac{1-\gamma}{\gamma-2}}\left(\frac{1-\alpha^2}{4}\right)^{\frac{1}{2\gamma-4}}\nonumber
\\&&\qquad\qquad \times\Bigg{\{}
-\frac{1}{4}\Gamma\left(\frac{\alpha^2+1}{4},\frac{1-\alpha^2}{4}L_{W}(\eta_{+})\right)
+\frac{1}{\alpha^2-1}\left[\Gamma\left(\frac{\alpha^2+5}{4},\frac{1-\alpha^2}{4}L_{W}(\eta_{+})\right)
-\frac{1}{2}\Gamma\left(\frac{\alpha^2+1}{4}\right)\right]
\Bigg{\}}.
\end{eqnarray}
\subsection{First law of thermodynamics and phase structure}
We start this part of paper by calculating thermodynamic variables
to check the first law of black hole thermodynamics. We consider
cosmological constant as black hole pressure and its associated
conjugate as volume of black hole. As mentioned above, entropy of
black hole is related to its horizon area, so we can obtain the
thermodynamic volume of black hole as
\begin{equation}\label{vol}
V=\int 4S
dr_+=\frac{(\alpha^2+1)b^{\gamma}}{(\alpha^2+3)r_+^{\gamma-3}}\omega
\end{equation}
As we take cosmological constant as the black hole pressure, so
the ADM mass should be interpreted as enthalpy, $H\equiv M$ rather
than the internal energy \cite{enthalpy}, and it should be a
function of extensive quantities: entropy and charge, and
intensive quantities: pressure and nonlinear parameter. Indeed, in
the extended phase space, another thermodynamic variable is the
nonlinear parameter $\beta$, which its conjugate is defined as
\cite{MannBI}
\begin{equation}\label{B}
B=\left(\frac{\partial M}{\partial \beta}\right)_{S,Q,P}.
\end{equation}
Therefore, the first law takes the form
\begin{equation} \label{FL}
dM=TdS+UdQ+VdP+Bd\beta.
\end{equation}%
The conjugate of $\beta$ has the dimension of polarization per
unit volume and can interpret as vacuum polarization \cite{GW}.
Throughout this paper, we choose the unit in which, from
dimensional analysis, one can find $[\beta]=[m]=[b]=[q]$, and
$\alpha$ is a dimensionless parameter. We shall also investigate
the effects of both dilaton parameter $\alpha$ as well as the
nonlinear parameter $\beta$ on the critical behaviour and phase
structure of the nonlinear dilaton black holes.

According to definition (\ref{B}), the conjugate quantity of
nonlinear parameter for END black hole is given by
\begin{eqnarray}
&&B\mid_{\rm END}=\frac{b^\gamma
(\alpha^2+1)\omega}{8\pi}\Bigg{\{}\frac{4\beta b^\gamma
r_+^{3-2\gamma}}{\alpha^2-3}+q(\alpha^2+1)\left(\frac{\beta
b^\gamma}
{q}\right)^{\frac{1-\gamma}{\gamma-2}}L_{W}(\eta_+)^{\frac{3-2\gamma}{2\gamma-4}}\nonumber
\\
&&\times
\Bigg[\frac{L_W^2(\eta_+)}{\alpha^2+5}\digamma\left(\frac{\alpha^2+5}{4},\frac{\alpha^2+9}{4},\frac{\alpha^2-1}{4}L_W(\eta_+)\right)
-\frac{1}{\alpha^2-3}\digamma\left(\frac{\alpha^2-3}{4},\frac{\alpha^2+1}{4},\frac{\alpha^2-1}{4}L_W(\eta_+)\right)\Bigg]\Bigg{\}}.
\end{eqnarray}
In the linear regime where $\beta\rightarrow\infty $, the
conjugate of nonlinear parameter goes to zero. As an example, let
us expand $B$  for large $\beta$ for $\alpha=0,1$. We find
 \begin{eqnarray}\label{B-Aym-END}
 &&B\mid_{\rm END}= \left\{
 \begin{array}{ll}
 $$\frac{q^4 \omega}{80\pi  r^5\beta^3}-\frac{5 q^6\omega}{432\pi r^9 \beta^5}+O\left(\frac{1}{\beta^7}\right)~~~~~~~{\rm for  \   \   \alpha=0,}  \\
 $$
 &  \\
 $$\frac{q^4 \omega}{48\pi  r^3 b^2\beta^3}-\frac{q^6\omega}{48\pi r^5 b^4
 \beta^5}+O\left(\frac{1}{\beta^7}\right)
 ~~~~{\rm for \  \   \alpha=1.} $$&  \\  &
 \end{array} \right.
 \end{eqnarray}
One can calculate the pressure as
\begin{eqnarray}\label{press}
P=\frac{(\alpha^2+3)
b^\gamma}{(\alpha^2-3)r_+^\gamma}\frac{\Lambda}{8\pi},
\end{eqnarray}
which is in accordance with the result of \cite{Kamrani,Dayyani1}.
In the absence of dilaton field ($\alpha=0=\gamma$), the above
expression for pressure reduces to the pressure of RN-AdS black
holes in an extended phase spaces \cite{MannRN}. It is easy to
show that all conserved and thermodynamic quantities in this
theory satisfy the first law of black hole thermodynamics
(\ref{FL}). Using scaling (dimensional) argument, the
corresponding Smarr formula per unit volume $\omega$ can be
written as
\begin{equation}\label{smar}
M=\frac{2}{\alpha^2+1}TS+\frac{\alpha^2-1}{\alpha^2+1}(2VP+\beta
B)+UQ.
\end{equation}
One can easily check that in limiting case $\alpha=0$, this
relation is exactly Smarr formula of \cite{Hendi1}, while in case
of linear Maxwell electrodynamics, it reduce to Smarr relation of
RN-AdS black hole \cite{MannRN}.
\subsection{Equation of state}
The critical point can be obtained by solving the following
equations
\begin{equation}\label{critical point}
\frac{\partial P}{\partial v}\Big{|}_{T_c}=0, \qquad\qquad
\frac{\partial^2 P}{\partial v^2}\Big|_{T_c}=0.
\end{equation}
In order to obtain the critical point, we should introduce the
equation of state  $P=P(V,T)$ by helping Eqs. (\ref{TemEND}) and
(\ref{press}). It is a matter of calculation to show
\begin{eqnarray}\label{eqstatEND1}
P\mid_{\rm END}=\frac{\Gamma
T}{r_+}+\frac{\Gamma(\alpha^2+1)}{2\pi}\Bigg[\frac{kr_+^{\gamma-2}}{2(\alpha^2-1)b^\gamma}+\frac{\beta^2
b^\gamma}{r_+^\gamma}+\frac{q
\beta}{r_+^2}\Big(\sqrt{L_W(\eta_+)}-\frac{1}{\sqrt{L_W(\eta_+)}}\Big)\bigg],
\end{eqnarray}
where we have defined
\begin{eqnarray}
\Gamma=\frac{(3+\alpha^2)}{2(3-\alpha^2)(\alpha^2+1)}.
\end{eqnarray}
Note that Eq. (\ref{eqstatEND1}) does not depend on the volume
explicitly. However, if one pay attention to relation (\ref{vol}),
one see that the volume is a function of $r_+$. Thus, we can
rewrite relation (\ref{eqstatEND1}) as
\begin{equation}\label{eqstatEND2}
P\mid_{\rm
END}=\frac{T}{v}-\frac{k(\alpha^2+3)}{8\pi(\alpha^2-3)(\alpha^2-1)}\frac{(v\Gamma)^{\gamma-2}}{b^\gamma}
-\frac{\beta^2(\alpha^2+3)}{4\pi(\alpha^2-3)}\left(\frac{b}{v\Gamma}\right)^\gamma-\frac{q\beta
(\alpha^2+3)}{4\pi(\alpha^2-3)v^2\Gamma^2}\bigg(\sqrt{L_W(\eta^\prime)}-\frac{1}{\sqrt{L_W(\eta^\prime)}}\bigg),
 \end{equation}
where
\begin{equation}
\eta^\prime=\left(\frac{q~ (v\Gamma)^{\gamma-2}}{\beta~ b^\gamma
}\right)^2.
\end{equation}
It is interesting to study dimensional analysis of Eq.
(\ref{eqstatEND2}). Following \cite{MannRN}, we can write physical
pressure and temperature as
 \begin{equation}\label{press1}
\mathcal{P}=\frac{\hbar c}{l_{p}^{2}}P,\qquad
\mathcal{T}=\frac{\hbar c}{\kappa }T,
\end{equation}
where $l_{p}=\sqrt{\hbar G/c^{3}}$ is the Plank length, $\kappa $,
$\hbar$ and $c$ are the Boltzmann constant, Dirac constant and the
speed of light, respectively. Inserting Eq. (\ref{press1}) in Eq.
(\ref{eqstatEND1}), we can define specific volume as
\begin{equation}
 v=\frac{l_p^2 r_+}{\Gamma}.
\end{equation}
Hereafter, we set $\hbar=c=G=l_p=1$, for simplicity. In order to
find critical volume $v_c$, critical temperature $T_c$ and
critical pressure $P_c$ , we should solve Eq. (\ref{critical
point}). However, due to the complexity of equation of state, we
consider the large $\beta$ limit of Eq. (\ref{eqstatEND2}). It is
easy to show that
\begin{equation}
P_{\rm
END}\Big|_{\beta\rightarrow\infty}=\frac{T}{v}-\frac{(\alpha^2+3)}{8\pi(\alpha^2-3)b^\gamma}\Bigg{\{}\frac{k(v\Gamma)^{\gamma-2}}{
(\alpha^2-1)}+q^2(v\Gamma)^{\gamma-4}-\frac{q^4(v\Gamma)^{3\gamma-8}}{4
b^{2\gamma}\beta^2}\Bigg{\}}+O\left(\frac{1}{\beta^4}\right)
\end{equation}
Considering the large $\beta$ limit, we can obtain the properties
of the critical point as
\begin{eqnarray}\label{critlargebeta}
&&v_c=\frac{\Pi}{\Gamma}-\frac{q^4(\alpha^2+4)(\alpha^2+7)}{8\Gamma
k\beta^2b^{2\gamma}}\Pi^{(2\gamma-5)}+O\left(\frac{1}{\beta^4}\right),\nonumber\\\nonumber\\*[.1cm]&&
P_c=\frac{q^2(\alpha^2+3)^2}{8\pi(\alpha^2-3)b^\gamma}\Pi^{\gamma-4}+\frac{q^4(\alpha^2+7)}{16\pi\beta^2b^{3\gamma}
}\Gamma
\Pi^{3\gamma-8}+O\left(\frac{1}{\beta^4}\right),\nonumber\\\nonumber\\*[.1cm]&&
 T_c=-\frac{k(\alpha^2+1)}{\pi(\alpha^2-1)(\alpha^2+3)b^\gamma}\Pi^{\gamma-1}+
 \frac{q^4(\alpha^2+4)}{8\pi\beta^2
 b^{3\gamma}}\Pi^{3\gamma-7}+O\left(\frac{1}{\beta^4}\right),
\end{eqnarray}
where
\begin{equation}
\Pi=\sqrt{\frac{q^2}{k}\left(\alpha^2+2\right)(\alpha^2+3)}.
\end{equation}
Let us note that Eq. (\ref{critlargebeta}) is similar to the
corresponding one in Born-Infeld-dilaton (BID) black holes
\cite{Dayyani1}. This is an expected result since for large
$\beta$  the equation of state of END and BID is exactly the same.
One can find that Eq. (\ref{critlargebeta}) follow the interesting
relation
\begin{equation}\label{universal ratio}
\rho
_{c}=\frac{P_{c}v_{c}}{T_{c}}=-\frac{(\alpha^2+3)(\alpha^2-1)}{4(\alpha^2+2)}\Bigg{\{}1-\frac{q^2}{2(\alpha^2+2)\beta^2
b^{2\gamma}}
\Pi^{2\gamma-4}\Bigg{\}}+O\left(\frac{1}{\beta^4}\right),
\end{equation}
In the absence of dilaton field ($\alpha=0$) and considering
linear electrodynamics where $\beta\rightarrow \infty$, we arrive
at $\rho_c={3}/{8}$, which is a universal value for Van der Waals
fluid. This implies that the critical behavior of this type of
black holes resembles the Van der Waals gas \cite{MannRN}.

\begin{figure}
\centering \subfigure[$\alpha=0$, $\beta=2$, $q=b=1$ and
$k=1$]{\includegraphics[scale=0.4]{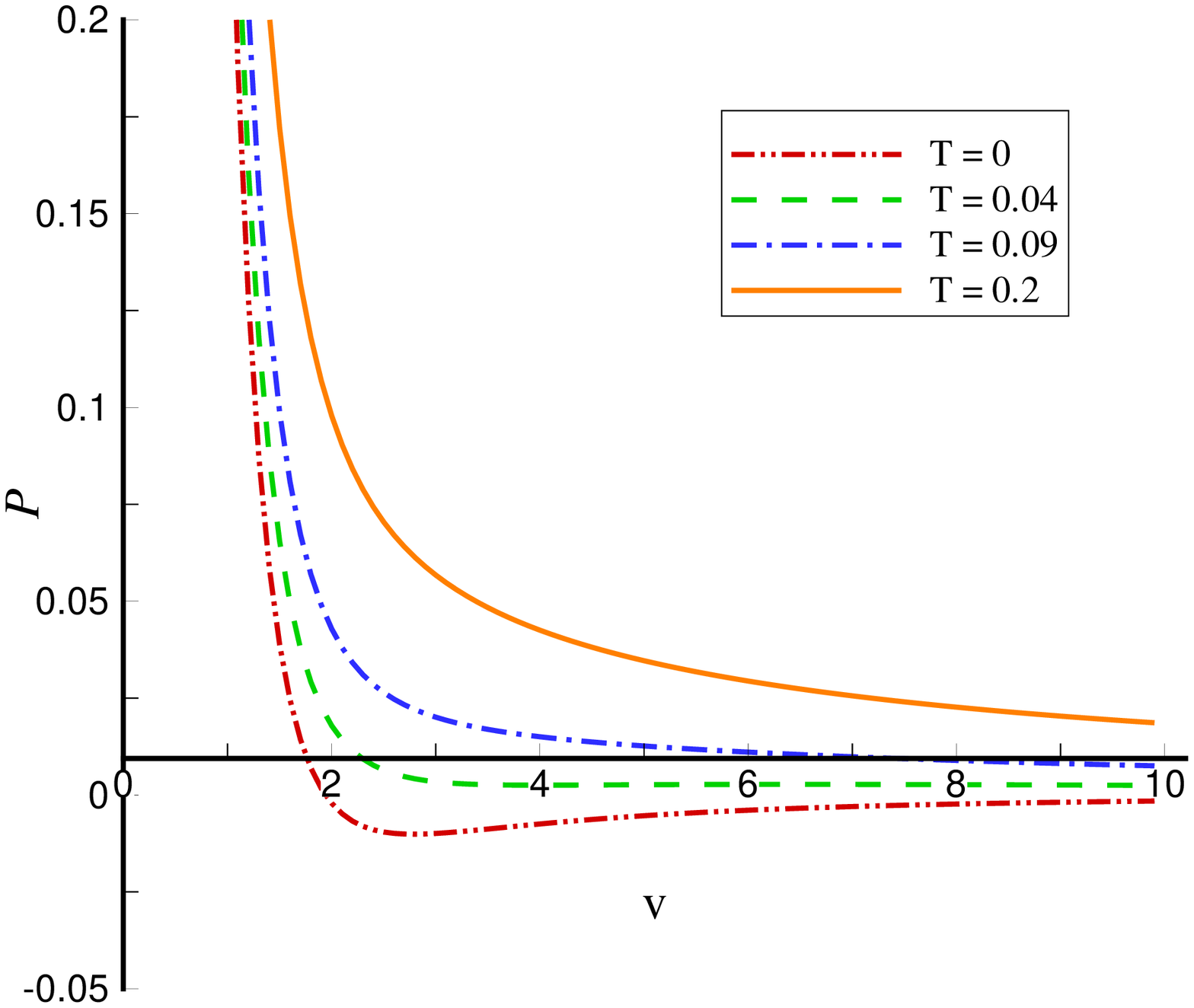}\label{fig1a}}
 \hspace*{.1cm} \subfigure[$\alpha=0.5$, $\beta=2$, $q=b=1$ and $k=1$
 ]{\includegraphics[scale=0.4]{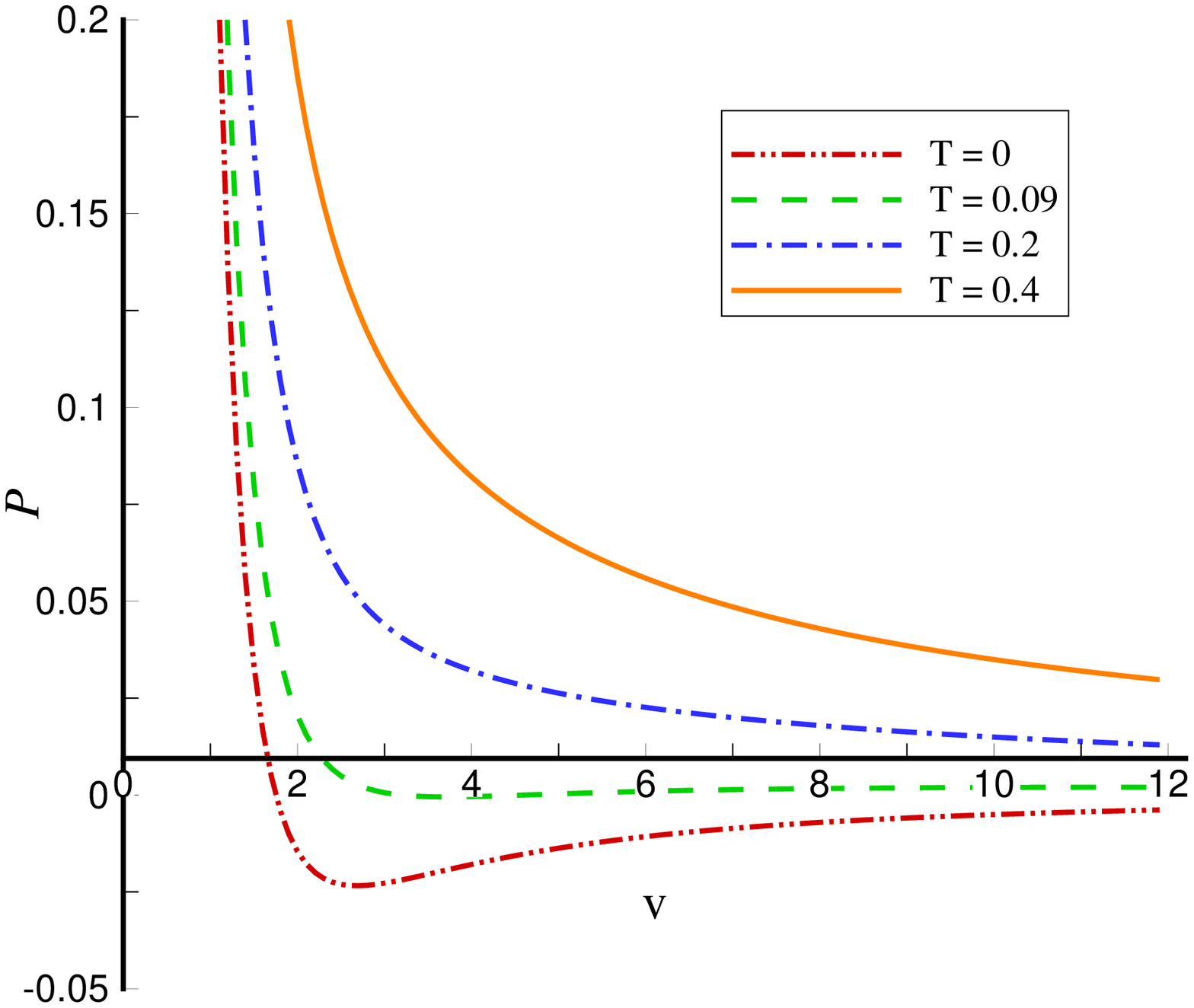}\label{fig1b}}\caption{$P-v$ diagram of END black holes.}\label{fig1}
 \end{figure}

To summarize, our solution can face with a phase transition when
temperature is below its critical value. One may predict this
behavior by considering isothermal $P-v$ diagram. It is expected
that $P-v$ diagram for our solution and Wan der Walls gas have
similar behaviour. In Fig. \ref{fig1} we have plotted the
behaviour of $P$ in terms of $v$. From these figures we see that,
in the absence/presence of dilaton field, the nonlinear black hole
resemble the  Van der Waals fluid behavior.
 \subsection{Gibbs free energy}
Another important approach to determine the critical behavior of a
system refers to study its thermodynamic potential. In the
canonical ensemble and extended phase space, thermodynamic
potential closely associates with the Gibbs free energy $G=M-TS$.
It is a matter of calculation to show that
 \begin{eqnarray}
G_{\rm
END}&=&\frac{\omega(\alpha^2+1)b^\gamma}{8\pi}\Bigg{\{}\frac{k~r_{+}}{2(\alpha^2+1)b^\gamma}
+\frac{(\Lambda+2\beta^2)(\alpha^2-1)b^\gamma}{2(\alpha^2-3)r_{+}^{2\gamma-3}}+2q\beta\left(\frac{q}{\beta
b^\gamma}\right)^{\frac{\gamma-1}{\gamma-2}}L_W(\eta_+)^{-\frac{2\gamma-3}{2\gamma-4}}
\nonumber\\&&\times\left[\frac{L_W(\eta_+)^2}{\alpha^2+5}\digamma\left(\left[\frac{\alpha^2+5}{4}\right],
\left[\frac{\alpha^2+9}{4}\right],\frac{\alpha^2-1}{4}L_W(\eta_+)\right)-\frac{1}{\alpha^2-3}\digamma
\left(\left[\frac{\alpha^2-3}{4}\right],\left[\frac{\alpha^2+1}{4}\right],\frac{\alpha^2-1}{4}L_W(\eta_+)\right)\right]\nonumber\\&&-q\beta
r_{+}^{1-\gamma}\left[\frac{1}{\sqrt{L_W(\eta_+)}}-\sqrt{L_W(\eta_+)}\right]\Bigg{\}}.
\end{eqnarray}
Expanding for large $\beta$  in the absence of dilaton field
($\alpha=0$), we arrive at
\begin{equation}\label{expandGexp}
G_{\rm
END}\Big|_{\beta\rightarrow\infty}=\frac{k~r_{+}}{4}-\frac{2\pi
r_{+}^3P}{3}+\frac{3 q^2}{4 r_{+}}-\frac{7q^4}{80
r_{+}^5\beta^2}+O\left(\frac{1}{\beta^2}\right).
\end{equation}
This is nothing but the Gibbs free energy of RN-AdS black holes
with a nonlinear leading order correction term \cite{MannRN}. In
order to study the Gibss free energy, we plot Fig. \ref{fig3a}.
One can see swallow-tail behavior in this figure which indicates a
phase transition under a critical value of temperature.
\subsection{Critical exponents}\label{exponent}
Here we would like to study critical exponents for END case. For
this purpose, we first calculate the specific heat as
 \begin{equation}
 C_V=T\left(\frac{\partial S}{\partial T}\right)\Big|_V.
 \end{equation}
We also redefine Eq. (\ref{entropy}) as
\begin{equation}
 S=S(V,T)=\frac{b^{\gamma}\omega}{4}\left[\frac{(\alpha^2+3)
 V}{(\alpha^2+1)b^\gamma}\right]^{{(\gamma-2)}/{(\gamma-3)}}.
\end{equation}
It is clear that entropy does not depend on the temperature in
this relation, so $C_V=0$. This indicates that relative critical
exponent will be zero
\begin{equation}
C_V\propto\left(\frac{T}{T_c}-1\right)^{\alpha^\prime}\Rightarrow\alpha^\prime=0.
\end{equation}
In order to find other critical exponent we consider the following
definition
\begin{equation}
\tau=\frac{T}{T_c}\qquad\qquad
p=\frac{P}{P_c}\qquad\qquad\nu=\frac{V}{V_c}.
\end{equation}\label{critdef}
Thus, we find
\begin{eqnarray}\label{Pcritexp}
p\mid_{\rm
END}&=&\frac{1}{\rho_c}\frac{\tau}{\nu}-\frac{k(\alpha^2+3)}{8\pi
P_c(\alpha^2-3)(\alpha^2-1)}\frac{(v_c\nu\Gamma)^{\gamma-2}}{b^\gamma}
-\frac{\beta^2(\alpha^2+3)}{4\pi
P_c(\alpha^2-3)}\left(\frac{b}{v_c\nu\Gamma}\right)^\gamma\nonumber\\&&-\frac{q\beta
(\alpha^2+3)}{4\pi
P_c(\alpha^2-3)(v_c\nu\Gamma)^2}\bigg(\sqrt{L_W(\eta^{\prime\prime})}-\frac{1}{\sqrt{L_W(\eta^{\prime\prime})}}\bigg),
 \end{eqnarray}
 where
 \begin{equation}
 \eta^{\prime\prime}=\left(\frac{q~ (v_c\nu\Gamma)^{\gamma-2}}{\beta~ b^\gamma
 }\right)^2.
 \end{equation}
 Expanding for $\beta\rightarrow\infty$, yields
\begin{eqnarray}
p=\frac{1}{\rho_c}\frac{\tau}{\nu}+\frac{1}{4\pi
P_c}\Bigg{\{}\frac{k (\alpha^2+1)\Gamma
b^{-\gamma}}{(\alpha^2-1)(\nu\Gamma
v_c)^{2-\gamma}}-\frac{q^2(\alpha^2+3)(\nu\Gamma
v_c)^{2\gamma-4}}{2(\alpha^2-3)b^{\gamma}(\nu
\Gamma)^\gamma}\Bigg{\}}.
\end{eqnarray}
Since we would like to find critical exponent, we should consider
the close neighborhood of critical point, so we expand Eq.
(\ref{Pcritexp}) near the critical point. Considering $\tau=t+1$
and $\nu=(\omega+1)^{1/\epsilon}$ where
$\epsilon=(\alpha^2+3)/(\alpha^2+1)$, and taking into account
relation (\ref{Pcritexp}), we get
\begin{equation}\label{pexpcrit}
p=1+At-Bt\omega-C\omega^3+O(t\omega^2,\omega^4),
\end{equation}
where
\begin{equation}
A=\frac{1}{\rho_c}\qquad B=\frac{1}{\epsilon\rho_c}\qquad
C=\frac{2(\alpha^2+2)}{3(\alpha^2+1)^2\epsilon^3}-\frac{(\alpha^2+7)q^{2\gamma-2}}
{3(\alpha^2+3)^3\beta^2b^{2\gamma}
}\left(\frac{k}{(\alpha^2+3)(\alpha^2+2)}\right)^{2-\gamma}.
\end{equation}
According to the Maxwell's equal area law \cite{MannRN}, we get
\begin{eqnarray}\label{maxwellequal}
&&p=1+At-Bt\omega_l-C\omega_l^3=1+At-Bt\omega_s-C\omega_s^3\nonumber\\&&0=
-P_c\int\limits_{\omega_l}^{\omega_s}\omega(Bt+3C\omega^2)d\omega,
\end{eqnarray}
where we $\omega_l$ and $\omega_s$ refers to volume of large and
small black holes. The only non-trivial solution of Eq.
(\ref{maxwellequal}) is
\begin{equation}
\omega_l=\omega_s=2\sqrt{-\frac{B t}{C}}.
\end{equation}
The behavior of the order parameter near the critical point can be
found as
\begin{eqnarray}
\Xi=V_c(\omega_l-\omega_s)=2V_c\omega_l=2\sqrt{-\frac{B}{C}}t^{1/2}.
\end{eqnarray}
Therefore, the critical exponent associated with the order
parameter should be $\beta^\prime=\frac{1}{2}$ which coincides
with that in Van der Waals gas.  Isothermal compressibility near
the critical point can be obtained as
\begin{eqnarray}
\kappa_T=-\frac{1}{V}\frac{\partial V}{\partial
P}\Bigg|_T\propto-\frac{V_c}{B P_c}\frac{1}{t}.
\end{eqnarray}
Since $\kappa_T\propto t^{-\gamma^\prime}$, we have
$\gamma^\prime=1$ and as we expect near the critical point it
should diverge. The last critical exponent is $\delta^\prime$
which describes the relation between order parameter and ordering
field in the critical point, so we should set $t=0$ in Eq.
(\ref{pexpcrit}). We find
\begin{equation}
p-1=-C\omega^3\qquad\Longrightarrow\qquad\delta^\prime=3.
\end{equation}
It is important to note that all critical exponents in this theory
coincide with those of Van der Waals gas system.
%%%%%%%%%%%%%%%%%%%%%%%%%%%%%%%%%%%%%%%%%%%%%%%%%%%%%%%%%%%%%%%%%%%%%%%%%%%%%%%
\section{Critical behavior of LND black holes} \label{LND}
Now, we can repeat all above steps for LND electrodynamics and
consider the effect of this type of nonlinear electrodynamics on
the critical behaviour of the solutions. At first, we introduce
metric function and vector potential for this type of black holes
\cite{sara}
 \begin{equation}
 A_t^{\rm LND}=\frac{q}{r}~{}_3\digamma_2\left(\left[\frac{1}{2},1,\frac{\alpha^2+1}{4}\right],\left[2,\frac{\alpha^2+5}{4}\right],-\eta\right)
 \end{equation}
\begin{eqnarray}
f(r)_{\rm LND}&=&-k\frac{ {\alpha}^{2}+1}{ {\alpha}^{2}-1
}{b}^{-\gamma}{r}^{\gamma}-\frac{m}{r^{1-\gamma}}+\frac{(\Lambda-4\beta^2)
\left( {\alpha}^{2}+1 \right) ^{2}{b}^{\gamma}}{\alpha^{2}-3
}r^{2-\gamma}+\frac{8 \beta^2
 {(\alpha^2+1)^2}}{(\alpha^2-3)^2} b^\gamma r^{2-\gamma}
\nonumber\\
&&\times\Bigg{\{}1-{}_2\digamma_1\left(\left[-\frac{1}{2},\frac{\alpha^2-3}{4}\right],\left[\frac{\alpha^2+1}{4}\right],-\eta\right)
+
\frac{\alpha^2-3}{2}\left(\sqrt{1+\eta}-\ln\left(\frac{\eta}{2}\right)+\ln\left(-1+\sqrt{1+\eta}\right)\right)
\Bigg{\}}
\end{eqnarray}
where ${}_2\digamma_1$ and ${}_3\digamma_2$ is the hypergeometric
functions. In order to  study thermodynamics quantities, we first
find temperature as
\begin{eqnarray}\label{TemLND}
&&T_{+}^{\rm LND}=-2k \frac{r_+^{\gamma-1} b^{-\gamma}}{4\pi}-\frac{m(\alpha^2-3)}{4\pi (\alpha^2+1)}r_+^{\gamma-2}+\frac{8\beta^2(\alpha^2+1)}{4\pi (\alpha^2-3)}b^\gamma r^{1-\gamma}\left(1-\frac{1}{\sqrt{1+\eta_+}}\right)+\frac{8q^2}{4\pi(\alpha^2-3)}b^{-\gamma}r_+^{\gamma-3}\nonumber\\
&&\qquad\qquad\times\Bigg{\{}2 \times
{}_2\digamma_1\left(\left[\frac{1}{2},\frac{\alpha^2+1}{4}\right],\left[\frac{\alpha^2+5}{4}\right],-\eta_+\right)-\frac{\alpha^2+1}{\sqrt{1+\eta_+}}
\Bigg{\}}.
\end{eqnarray}
The entropy expression is the same as END case, because it does
not depend on electrodynamics and still obeys the area law.
Considering the definition of electric potential, one may obtain
$U$ as
\begin{eqnarray}\label{potLND}
U_{\rm
LND}=\frac{q}{r_+}~{}_3\digamma_2\left(\left[\frac{1}{2},1,\frac{\alpha^2+1}{4}\right],\left[2,\frac{\alpha^2+5}{4}\right],-\eta_+\right).
\end{eqnarray}
In order to verify the first law of thermodynamics, we should
calculate conjugate of nonlinear parameter for LND topological
black hole. We obtain
\begin{eqnarray}
B_{\rm LND}&=&\frac{(\alpha^2+1)\beta
b^{2\gamma}}{\pi(\alpha^2-3)r_+^{2\gamma-3}}\Bigg{\{}\sqrt{1+\eta_+}-\ln\left(\frac{\eta_+}{2}\right)+\ln(\sqrt{1+\eta_+}-1)-\frac{\eta_+}{2
\sqrt{1+\eta_+}}\left(1+\frac{1}{\sqrt{1+\eta_+}-1}\right)+\frac{2}{\alpha^2-3}\nonumber\\&&\times\left(1-{}_2\digamma_1\left(\left[-\frac{1}{2},\frac{\alpha^2-3}{4}\right],\left[\frac{\alpha^2+1}{4}\right],-\eta_+\right)\right)+\frac{\eta_+}{(\alpha^2+1)}~{}_2\digamma_1\left(\left[\frac{1}{2},\frac{\alpha^2+1}{4}\right],\left[\frac{\alpha^2+5}{4}\right],-\eta_+\right)
\Bigg{\}},
\end{eqnarray}
which its asymptotic behavior for $\beta\rightarrow\infty$ and
$\alpha=0,1$ can be obtained as
  \begin{eqnarray}
&& B_{\rm LND}\Big|_{\beta\rightarrow\infty,\alpha=0}=\frac{q^4
\omega}{160\pi
 r^5\beta^3}-\frac{ q^6\omega}{432\pi r^9 \beta^5}+O\left(\frac{1}{\beta^7}\right)\nonumber\\
&& B_{\rm LND}\Big|_{\beta\rightarrow\infty,\alpha=1}=\frac{q^4
\omega}{96\pi  r^3 b^2\beta^3}-\frac{q^6\omega}{240\pi r^5 b^4
\beta^5}+O\left(\frac{1}{\beta^7}\right).
  \end{eqnarray}
It is clear that this relation is similar to those given in Eq.
(\ref{B-Aym-END}). Definition of black holes thermodynamic volume
is related to the entropy and since the entropy expression does
not depend on the type of electrodynamics, so thermodynamics
volume is the same as given in Eq. (\ref{vol}). Also, as we
mentioned before, the pressure is related to the cosmological
constant, so for LND black holes, one can find that the pressure
is exactly the same as given in Eq. (\ref{press}). Finally, it is
a matter of calculation to check that all conserved and
thermodynamic quantities of LND black holes satisfy the first law
of black thermodynamics (\ref{FL}) as well as Smarr relation
(\ref{smar}).
\subsection{Equation of state}
\begin{figure}
\centering \subfigure[$\alpha=0.3$, $\beta=2$, $q=b=1$ and
$k=1$]{\includegraphics[scale=0.4]{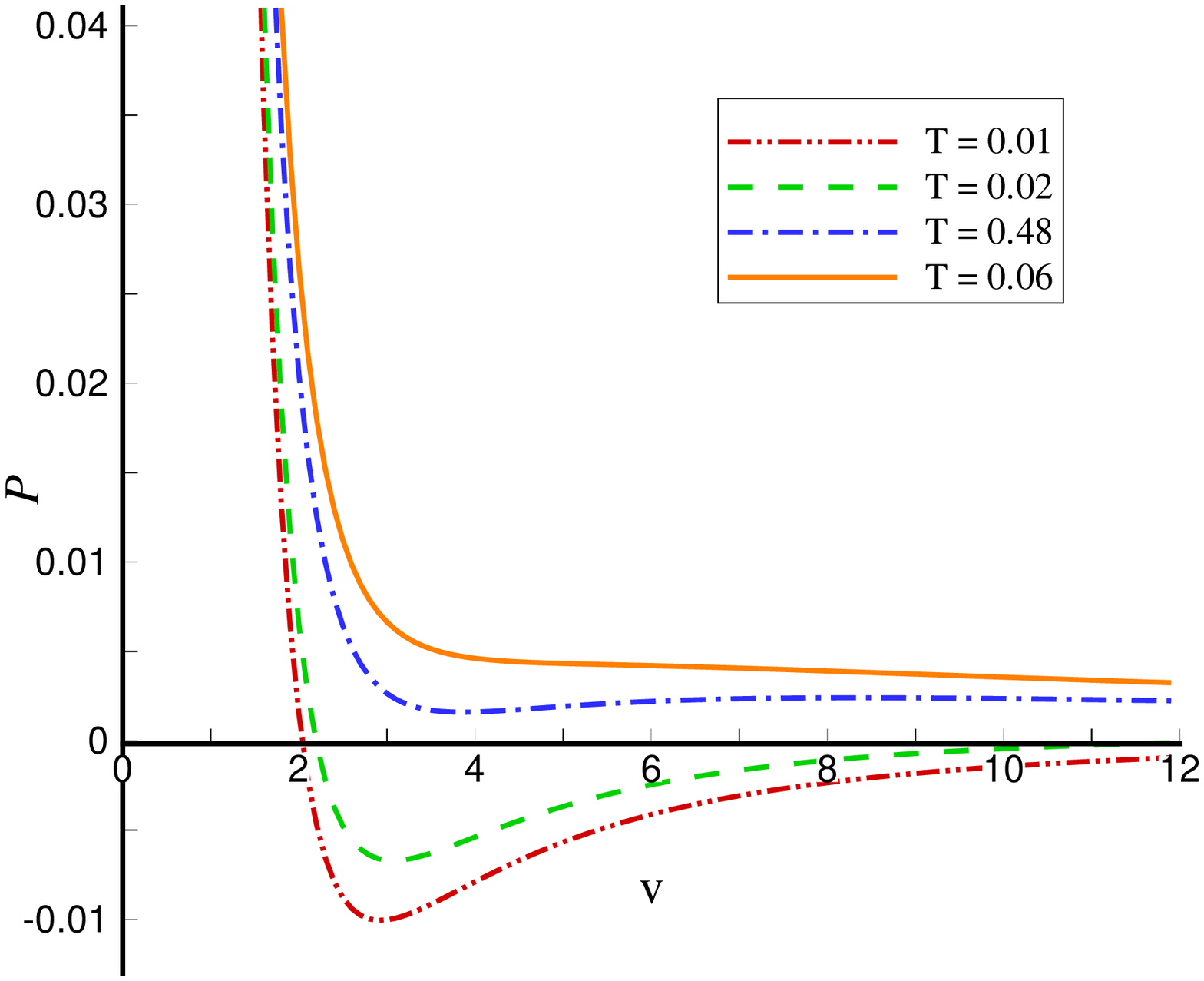}\label{fig2a}}
\hspace*{.1cm} \subfigure[$\alpha=\beta=0.2$, $q=b=1$ and $k=1$
]{\includegraphics[scale=0.4]{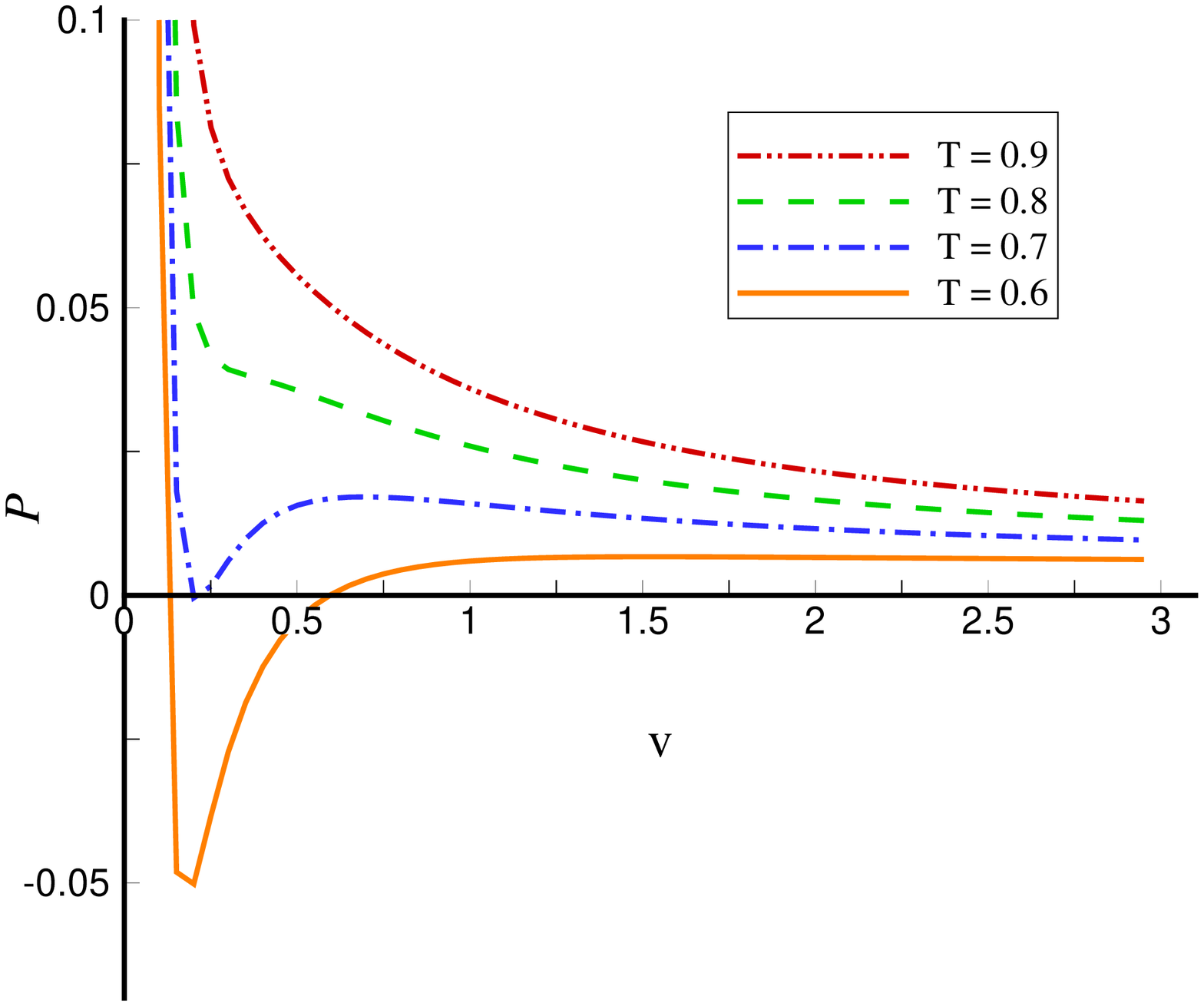}\label{fig2b}}\caption{$P-v$
diagram of LND black holes}\label{fig2}
\end{figure}

This section is devoted to study the critical behavior of black
hole in the presence of LND electrodynamics. In this regard, we
obtain equation of state at first
\begin{eqnarray}\label{plnd}
P_{\rm LND}&=&\frac{T}{v}+\frac{1}{4\pi v}\frac{k~(\alpha^2+1)}{b^\gamma
(\alpha^2-1)}(\Gamma v)^{\gamma-1}-\frac{4}{\pi v}\frac{q^2~(\Gamma v)^{\gamma-3}}{b^\gamma(\alpha^2-3)}{}_2\digamma_1
\Bigg(\left[\frac{1}{2},\frac{\alpha^2+1}{4}\right],\left[\frac{\alpha^2+5}{4}\right],-\eta^\prime\Bigg)+\frac{1}
{\pi v}\frac{\beta^2 b^\gamma(\alpha^2+1)}{(\Gamma v)^{\gamma-1}}\nonumber\\
&&\times\Bigg{\{}-\frac{2}{\alpha^2-3}{}_2\digamma_1\Bigg(\left[-\frac{1}{2},\frac{\alpha^2-3}{4}\right],
\left[\frac{\alpha^2+1}{4}\right],-\eta^\prime\Bigg)+\ln(\sqrt{1+\eta^\prime}-1)-\ln(\eta^\prime)+\ln(2)-1\Bigg{\}}\nonumber\\
&&+\frac{1}{\pi v}\frac{(\alpha^4-1)b^\gamma
}{(\alpha^2-3)\sqrt{1+\eta^\prime}}\Big[q^2 b^{-2\gamma}(\Gamma
v)^{2\gamma-4}+\beta^2\Big] (\Gamma v)^{1-\gamma}.
\end{eqnarray}
It is a general belief that one can predict a Van der Walls like
behavior for a thermodynamic system by studying its $P-v$
diagrams. According to Fig. \ref{fig2} we can observe that for
specific values of parameters, phase transition exist below a
critical temperature. It occurs for both large (Fig.\ref{fig2a})
and small (Fig. \ref{fig2b}) value of nonlinear parameter in the
presence of dilaton field.

One may find the properties of critical point by using Eq.
(\ref{plnd}). However, due to the complexity of this equation, it
is not easy to investigate the critical point for arbitrary
nonlinear parameter. Therefore, we consider the large $\beta$
limit of Eq. (\ref{plnd}),
\begin{equation}\label{plndinfinity}
 P_{\rm LND}\Big|_{\beta\rightarrow\infty}=\frac{T}{v}+\frac{(\alpha^2+1)}{4\pi v b^\gamma}\Bigg{\{}-\frac{k}{(\alpha^2-1)}
 (v\Gamma)^{\gamma-1}+q^2(v\Gamma)^{\gamma-3}-\frac{q^4 }{8
 b^{2\gamma}\beta^2}(v\Gamma)^{3\gamma-7}\Bigg{\}}.
 \end{equation}
In the absence of dilaton field ($\alpha=0$), the equation of
state of RN-AdS black holes in an extended phase space
\cite{MannRN} is recovered with a leading order nonlinear
correction term
\begin{equation}
P_{\rm
LND}\Big|_{\beta\rightarrow\infty}=\frac{T}{v}-\frac{k}{2\pi
v^2}+\frac{2q^2}{\pi v^4}-\frac{4 q^4}{\pi v^8\beta^2}.
\end{equation}
Therefore, for large $\beta$ limit, the critical point is obtained
as
 \begin{eqnarray}\label{critlargebetaEND}
 &&v_c=\frac{\Pi}{\Gamma}-\frac{5q^4(\alpha^2+4)(\alpha^2+7)}{16\Gamma k\beta^2b^{2\gamma}}\Pi^{2\gamma-5}\nonumber+O\left(\frac{1}{\beta^4}\right),\\\nonumber\\*[.1cm]
 &&P_c=\frac{q^2(\alpha^2+3)^2}{8\pi(\alpha^2-3)b^\gamma}\Pi^{\gamma-4}+\frac{5q^4(\alpha^2+7)}{32\pi \beta^2 b^{3\gamma}}\Gamma\Pi^{3\gamma-8}+O\left(\frac{1}{\beta^4}\right),\nonumber\\\nonumber\\*[.1cm]&&
 T_c=-\frac{k(\alpha^2+1)}{\pi(\alpha^2-1)(\alpha^2+3)b^\gamma}\Pi^{\gamma-1}+\frac{5q^4(\alpha^2+4)}{16\pi\beta^2 b^{3\gamma}}\Pi^{3\gamma-7}+O\left(\frac{1}{\beta^4}\right),\nonumber\\\nonumber\\*[.1cm]&&
 \rho_c=-\frac{(\alpha^2+3)(\alpha^2-1)}{4(\alpha^2+2)}\Bigg{\{}1-\frac{5q^2}{4(\alpha^2+2)\beta^2 b^{2\gamma}}
 \Pi^{2\gamma-4}\Bigg{\}}+O\left(\frac{1}{\beta^4}\right).
 \end{eqnarray}
It is important to note that all above relations reduce to those
of RN-AdS black holes in an extended phase space  \cite{MannRN}
provided $\alpha=0$ and $\beta\rightarrow\infty$. Comparing the
results obtained here with relation (\ref{critlargebeta}), one can
find that the critical point in the large $\beta$ expansion for
both electrodynamics are  similar and the same as those of BID
given in \cite{Dayyani1}. This is an expected result since in the
large $\beta$ limit, the Lagrangian of all of these theories have
similar expansion, namely
\begin{eqnarray}
&&L_{\rm BID}=L_{\rm END}=-F^2
e^{-2\alpha\phi(r)}+\frac{F^4}{8\beta^2}e^{-6\alpha\phi(r)}+O\left(\frac{1}{\beta^4}\right),\nonumber\\&&
\qquad~~~~~ L_{\rm END}=-F^2
e^{-2\alpha\phi(r)}+\frac{F^4}{16\beta^2}e^{-6\alpha\phi(r)}+O\left(\frac{1}{\beta^4}\right).
\end{eqnarray}
Thus for large $\beta$ the equation of state and the critical
point properties of  BID, END and LND electrodynamics are the
same.
\subsection{Gibbs free energy}
\begin{figure}
 \centering \subfigure[END black holes]{\includegraphics[scale=0.4]{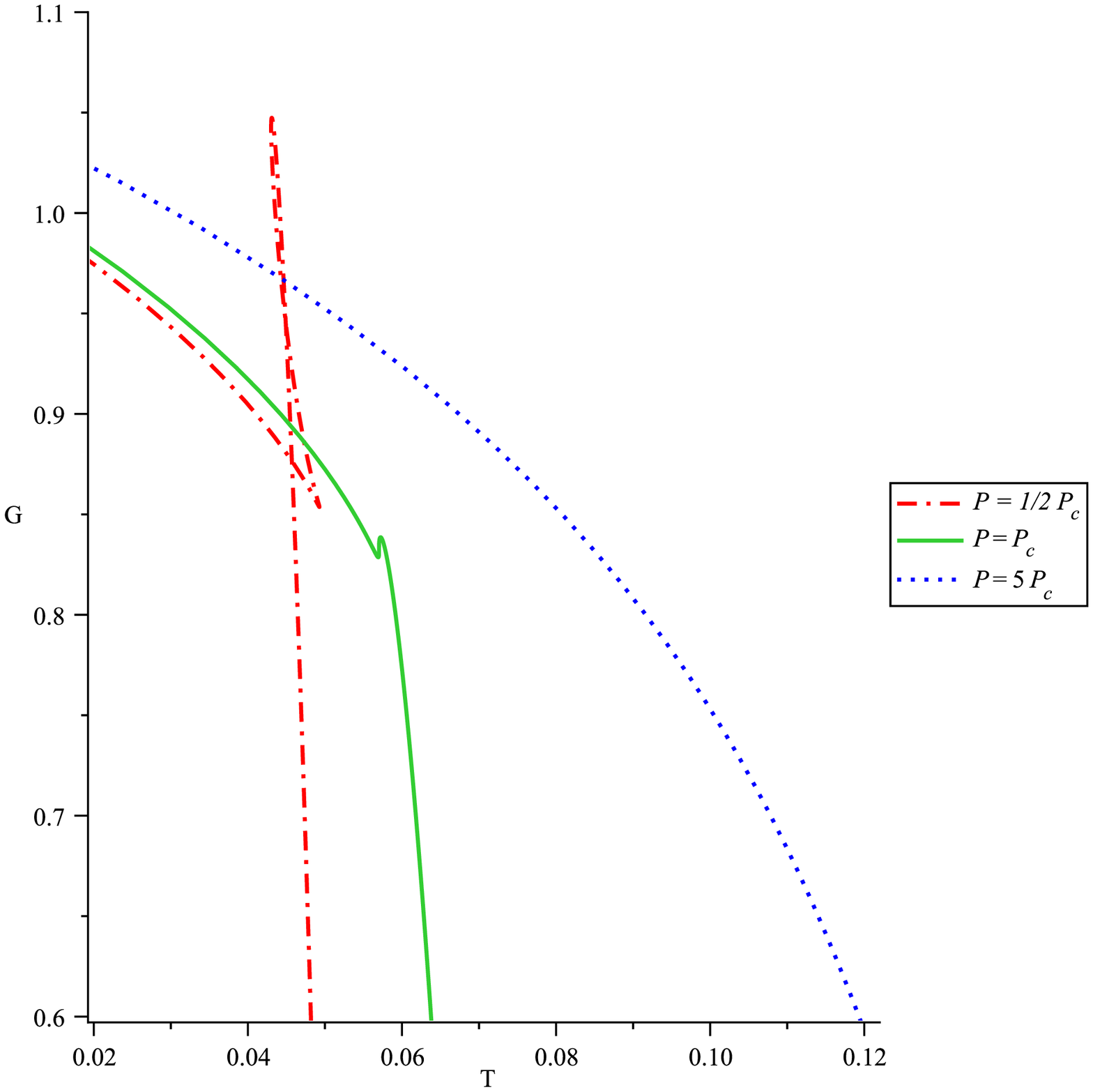}\label{fig3a}}
 \hspace*{.1cm} \subfigure[LND black holes]{\includegraphics[scale=0.5]{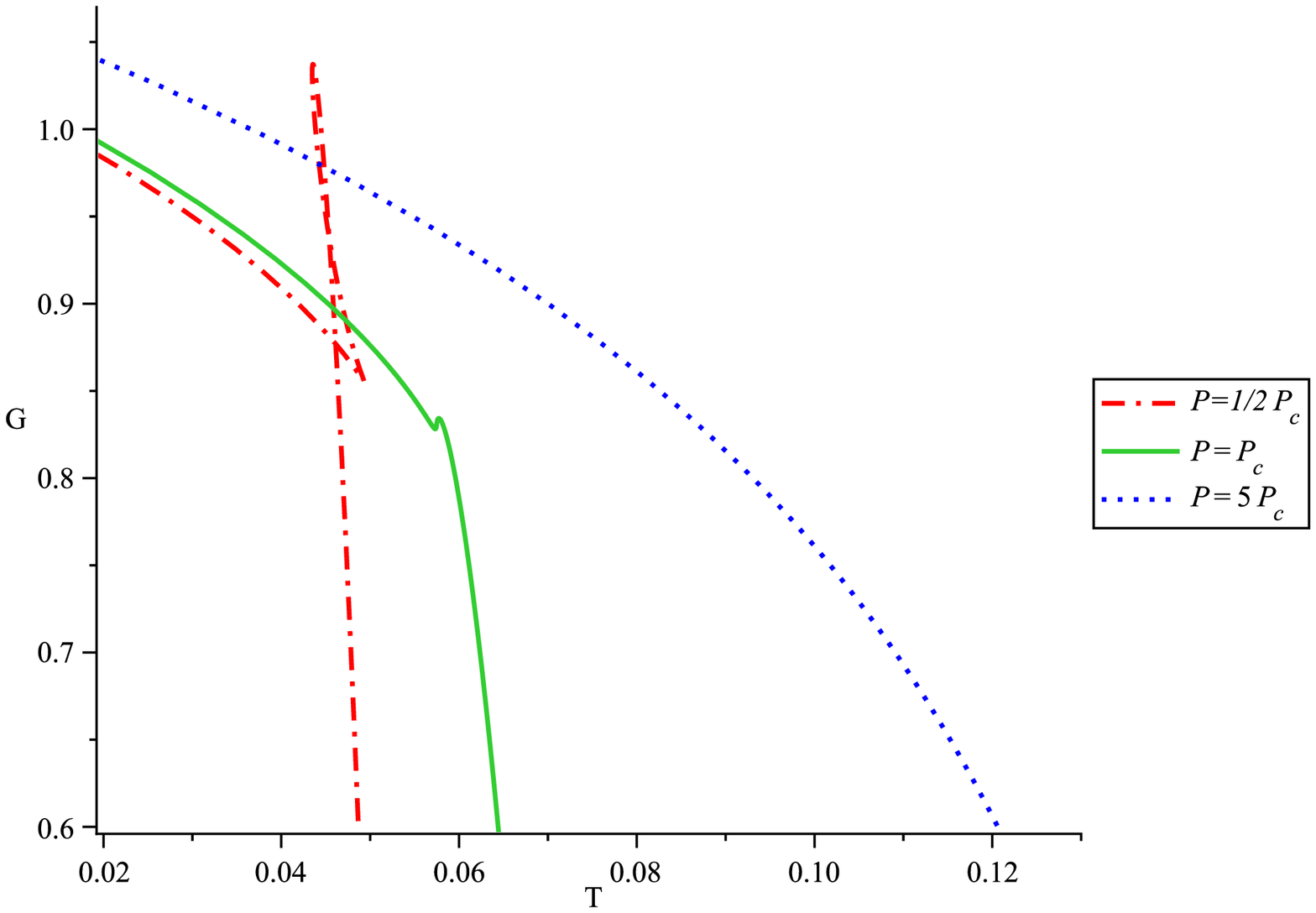}\label{fig3b}}\caption{Gibbs free energy versus $T$  for $\beta=q=b=k=1$ and $\alpha=0.3$.}\label{fig3}
 \end{figure}

Next, we study Gibbs free energy for LND black holes to
characterize phase transition in the system. It is a matter of
calculation to show that the Gibbs free energy of LND black holes
is given by
\begin{eqnarray}
G_{\rm
LND}&=&\frac{k~r_+\omega}{16\pi}+\frac{(\alpha^2+1)r^{3-2\gamma}\omega}{4\pi
(\alpha^2-3)}\Bigg{\{}\frac{2\pi
P(\alpha^2-3)(\alpha^2-1)(b~r_{+})^\gamma}{(\alpha^2+3)}+\frac{2q^2
r_{+}^{2\gamma-4}}{\sqrt{1+\eta_+}}\nonumber\\&&+b^{2\gamma}
\beta^2\Bigg[(\alpha^2-1)\ln\left(\frac{2\sqrt{1+\eta_+}}{\eta_+}-\frac{2}{\eta_+}\right)+\frac{(\alpha^2-1)\eta_++\alpha^2+1}{\sqrt{1+\eta_+}}-\frac{\alpha^4-4\alpha^2-1}{(\alpha^2-3)}\Bigg]-\frac{2(\alpha^2-1)\beta^2
b^{2\gamma}}{(\alpha^2-3)}\nonumber\\&&\times{}_2\digamma_1\Bigg(\left[-\frac{1}{2},\frac{(\alpha^2-3)}{4}\right],\left[\frac{(\alpha^2+1)}{4}\right],-\eta_+\Bigg)-\frac{4q^2
r_+^{2\gamma-4}}{(\alpha^2+1)}{}_2\digamma_1\Bigg(\left[\frac{1}{2},\frac{(\alpha^2+1)}{4}\right],\left[\frac{(\alpha^2+5)}{4}\right],-\eta_+\Bigg)\Bigg{\}}.
\end{eqnarray}
Note that if we expand this relation for large nonlinear parameter
$\beta$, we restore the result of Eq. (\ref{expandGexp}). We have
plotted the behavior of Gibbs free energy in term of temperature
in Fig. \ref{fig3b}. one can observes swallow-tail behavior in
this figure when pressure is smaller that its critical value. This
implies that the system experiments a phase transition.
\subsection{Critical exponents}
Next, we are going to obtain the critical exponent of LND black
holes. As we mentioned before, the entropy is equal in both
theories, so $C_v$ is equal too, and $\alpha^\prime =0$ like BID
and END theories. In order to calculate other critical exponent we
should follow the approach given in subsection \ref{exponent}. To
this end, we compute the equation of state near the  critical
point for LND theories
\begin{equation}\label{plndcrit}
p=1+At-Bt\omega-C^\prime\omega^3+O(t\omega^2,\omega^4),
\end{equation}
where
\begin{eqnarray}
C^\prime&=&\frac{2(\alpha^2+2)}{3(\alpha^2+3)^2\epsilon}-\frac{5(\alpha^2+7)q^{2\gamma-2}}{6(\alpha^2+3)^3\beta^2b^{2\gamma}
}\left(\frac{k}{(\alpha^2+3)(\alpha^2+2)}\right)^{2-\gamma}.
\end{eqnarray}
It is clear that the form of the above relation is similar to
relation (\ref{pexpcrit}), so as one expects all remind critical
exponent will be the same as in the case of END theory.
 %%%%%%%%%%%%%%%%%%%%%%%%%%%%%%%%%%%%%%%%%%%%%%%%%%%%%%%%%%%%%%%%%%%
\section{Effects of nonlinear gauge field}\label{Effect}
Although, we have calculated the critical quantities in the limit
of large $\beta$ where the nonlinearity of the theory is small.
However, it is clear from the $P-v$ and Gibbs diagrams that there
is a similar phase transition in the limit of small $\beta$ where
the nonlinearity of the theory is large. In the limit of small
$\beta$ it is nearly impossible to calculate analytically the
critical quantities. Also, in the presence of the dilaton field,
it will be very difficult to calculate them even numerically. For
some numeric calculations (in the absence of dilaton field) one
may see \cite{Hendi1}.

\begin{figure}
\centering \subfigure[  $T_{c}$ versus $\beta$
]{\includegraphics[scale=0.4]{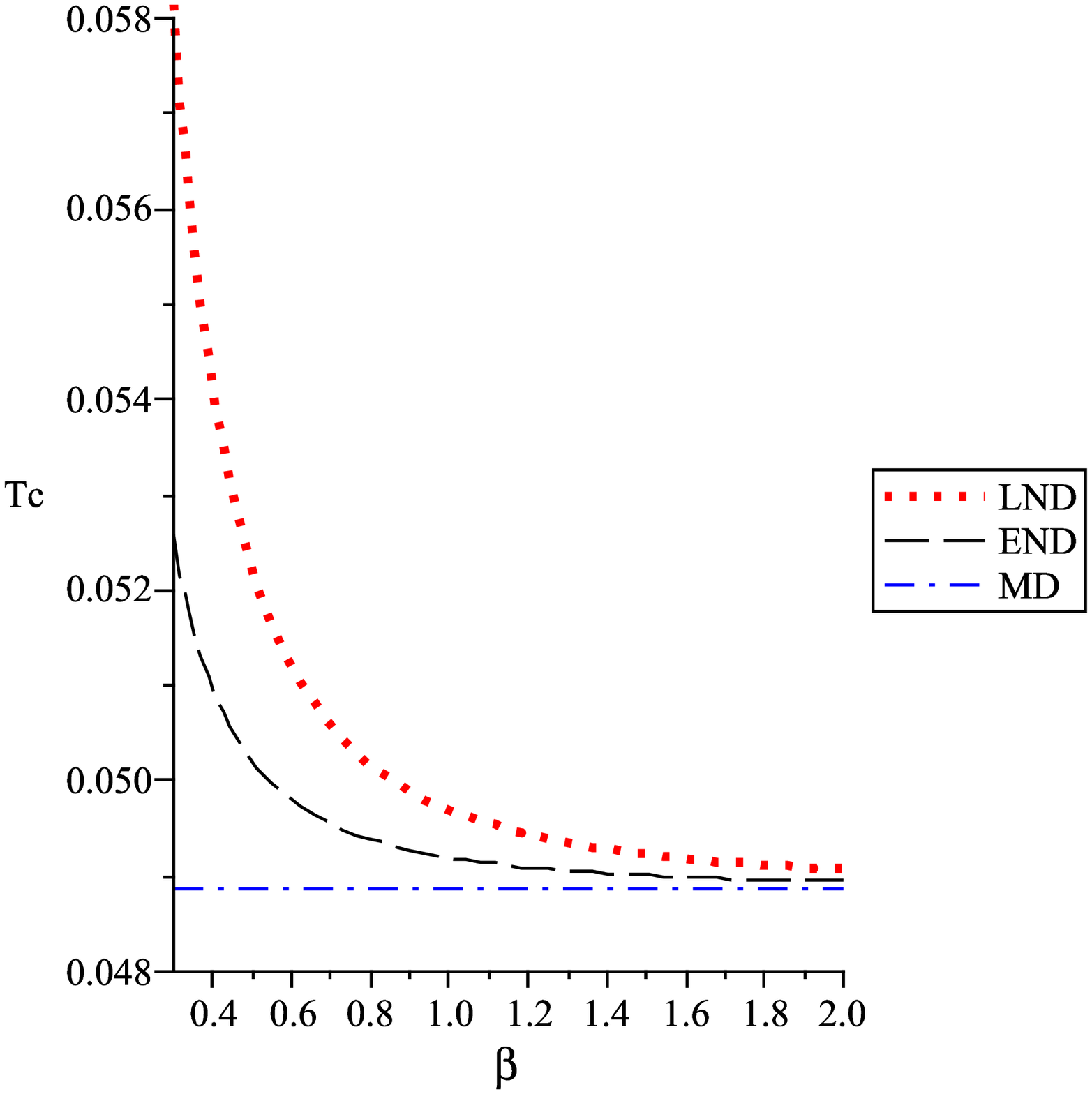}\label{fig8a}} \hspace*{.1cm}
\subfigure[   $P_{c}$ versus $\beta$
]{\includegraphics[scale=0.4]{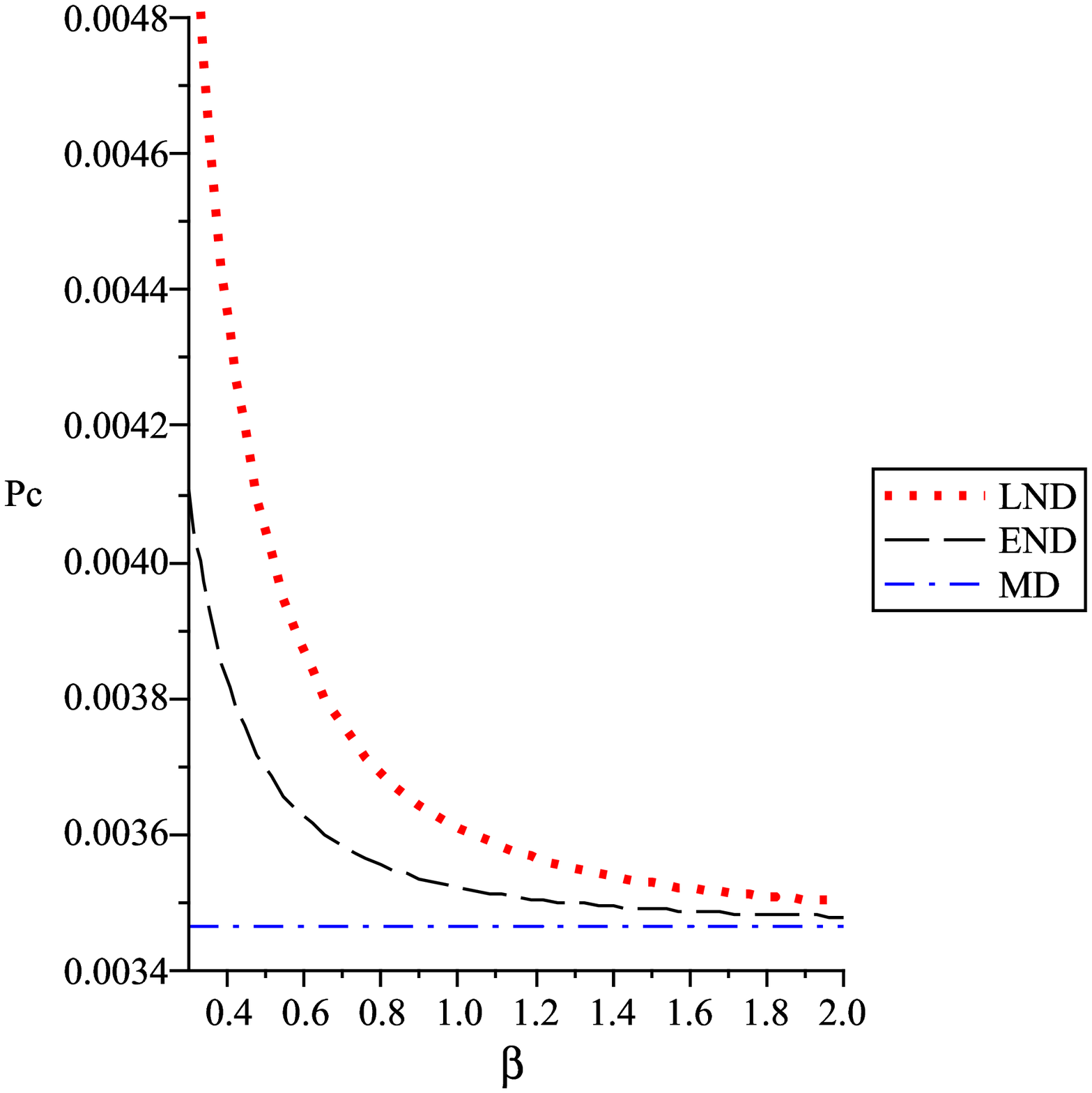}\label{fig8b}}\caption{Critical
quantities of dilaton black holes. Here we have set $q=b=k=1$ and
$\alpha=0.2$. }\label{fig8}
\end{figure}

\begin{figure}
\centering \subfigure[
$\beta=0.25$]{\includegraphics[scale=0.4]{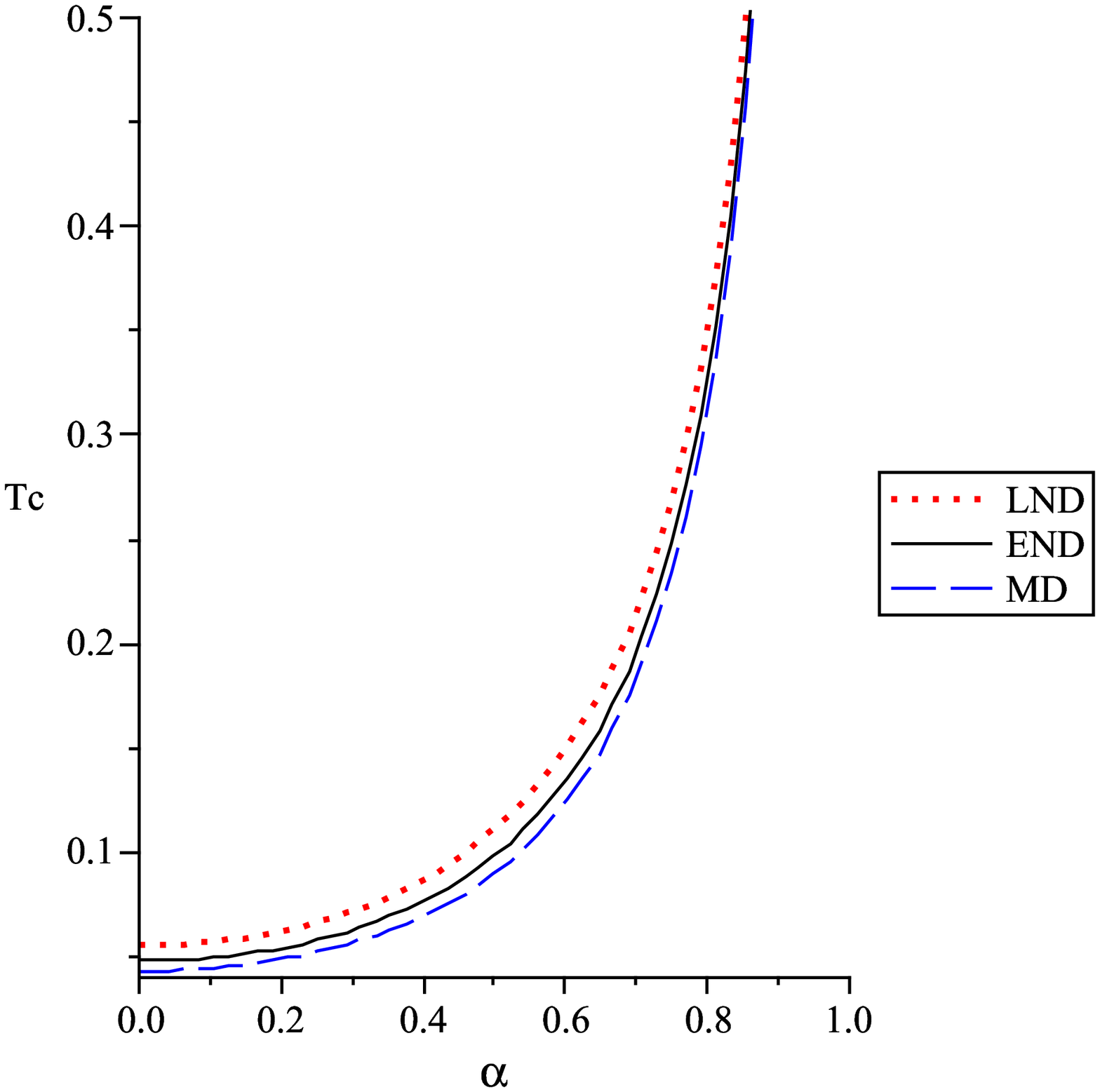}\label{fig9a}}
\hspace*{.1cm} \subfigure[ $\beta=0.1$
]{\includegraphics[scale=0.4]{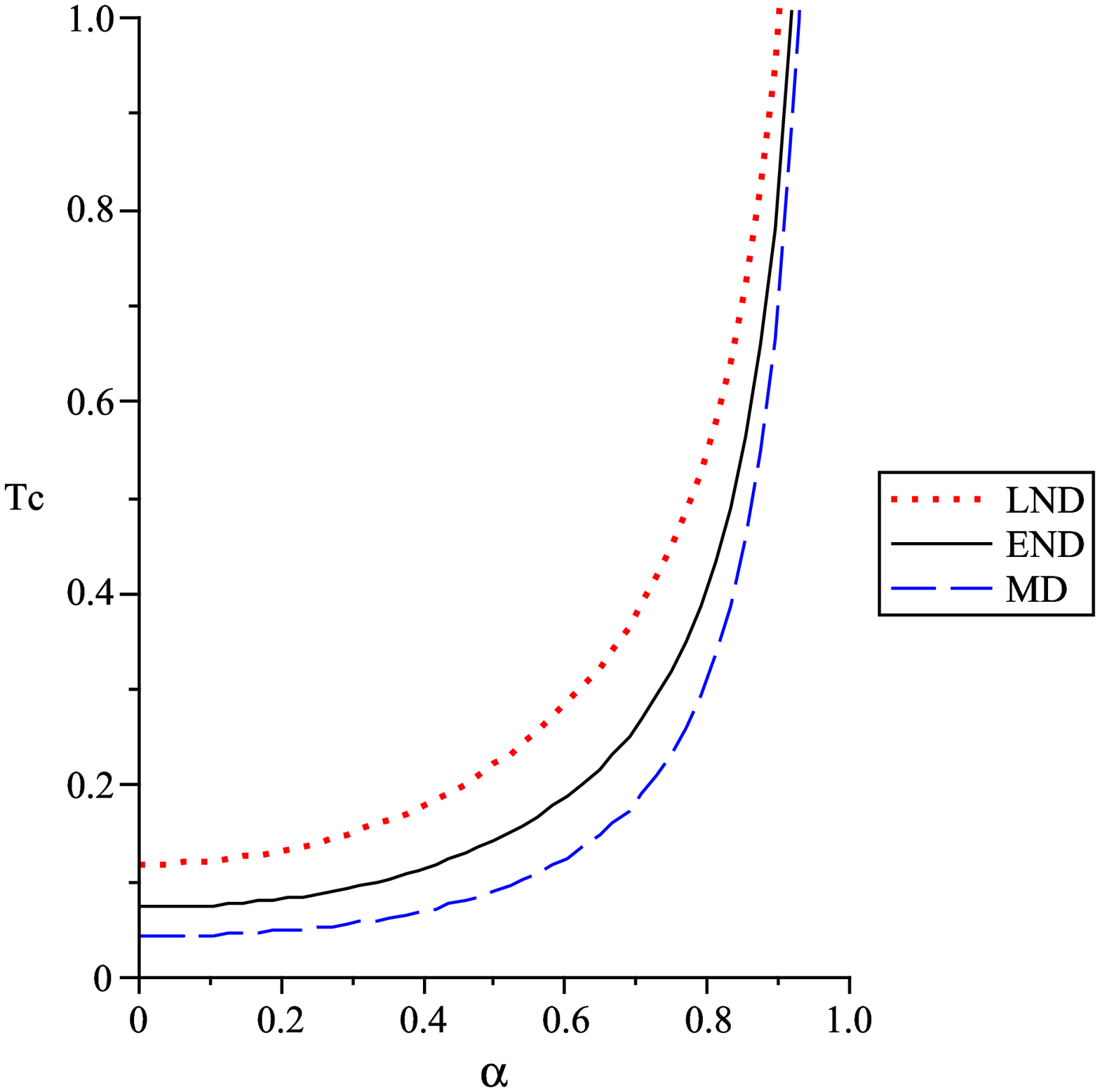}\label{fig9b}}\caption{$T_{c}$
versus $\alpha$  of dilaton black holes with different
electrodynamics. Here we have taken $q=b=k=1$.}\label{fig9}
\end{figure}

\begin{figure}
\centering \subfigure[
$\beta=0.5$]{\includegraphics[scale=0.4]{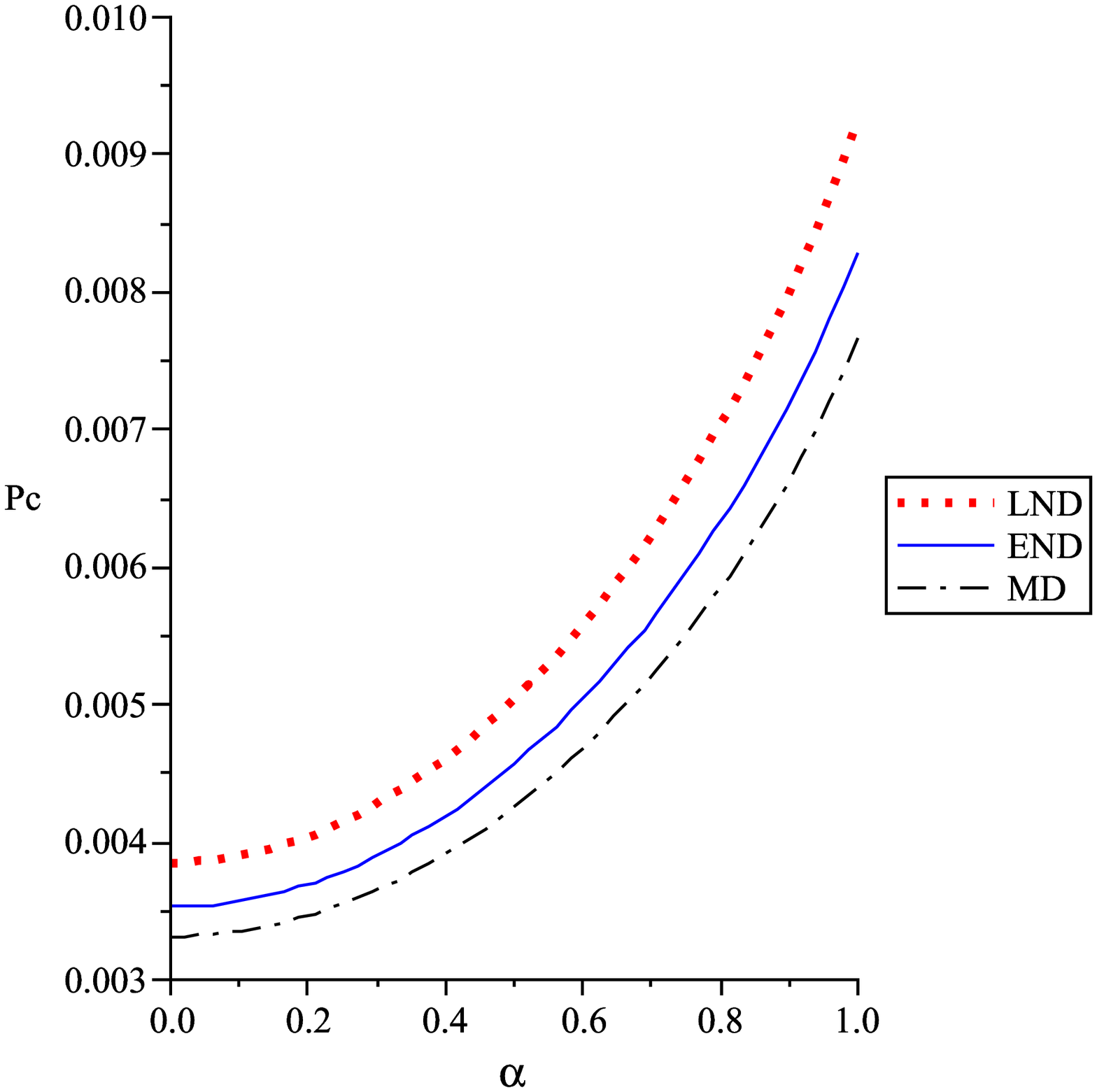}\label{fig10a}}
\hspace*{.1cm} \subfigure[ $\beta=0.1$
]{\includegraphics[scale=0.4]{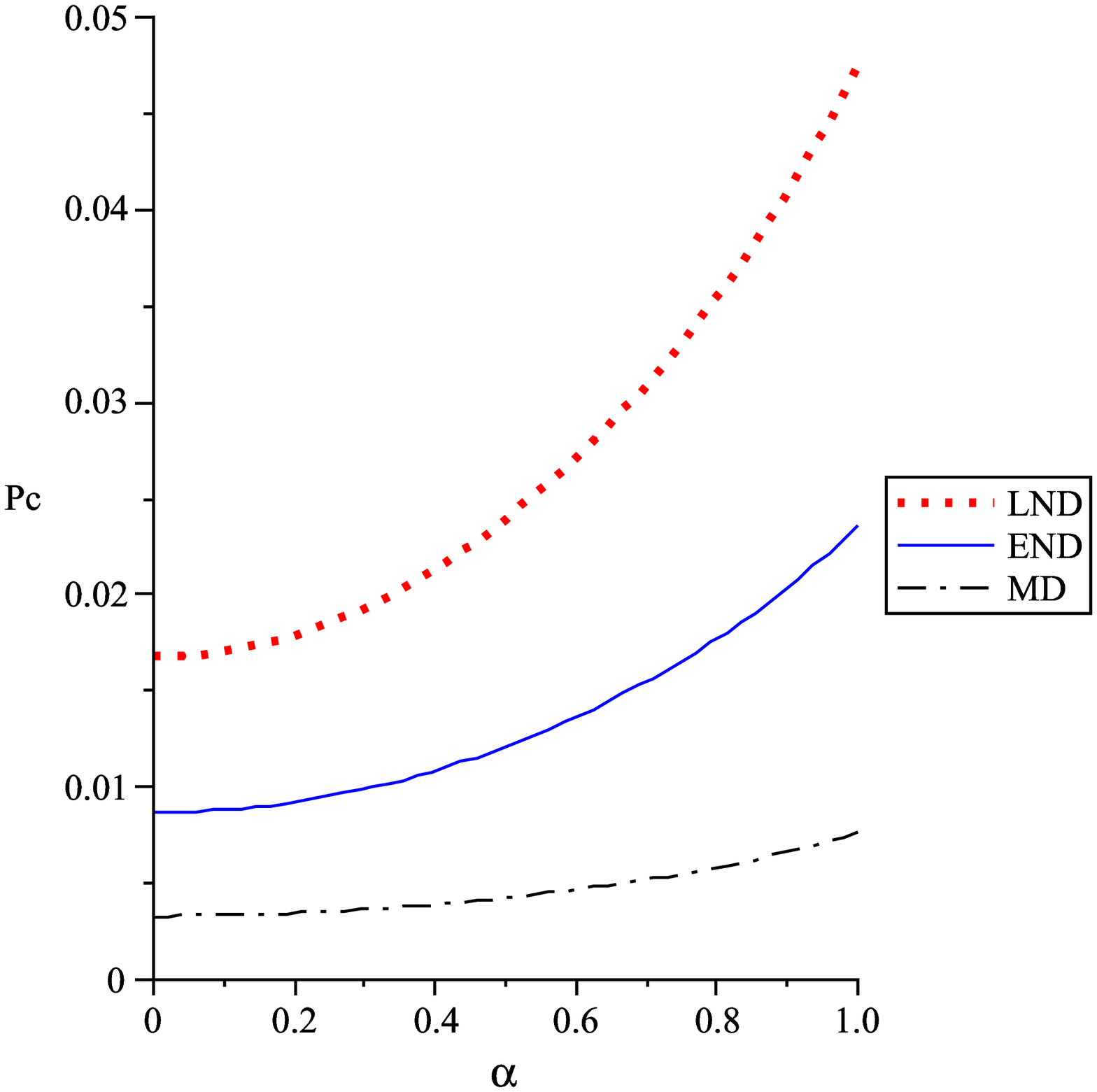}\label{fig10b}}\caption{$P_{c}$
versus $\alpha$ of black holes in different electrodynamics with
$q=b=k=1$}\label{fig10}
\end{figure}

A close look at the critical temperature in both END and LND given
in Eqs. (\ref{critlargebeta})  and (\ref{critlargebetaEND}), show
that the presence of the nonlinear field makes the critical
temperature larger and it will increase with decreasing $\beta$.
One may observe that the increasing in $T_{c}$ and  $p_{c}$ in LND
is stronger than END. In Fig. \ref{fig8} we have plotted critical
quantities $T_{c}$ and  $p_{c}$ of LND, END and Maxwell-dilaton
(MD) theory in terms of the nonlinear parameter $\beta$ and show
that they will go to a same value in the large limit of $\beta$
where the effects of nonlinearity disappears. Clearly, the linear
MD theory is independent of the nonlinear parameter $\beta$, as
can be seen from Fig. \ref{fig8}. It is notable to mention that
critical quantities in LND are the same as those in END for large
$\beta$. However, for small $\beta$ (nonlinear regime), their
behaviour is quit different. The behavior of the critical
temperature in term of $\alpha$ is shown in Fig \ref{fig9}, for
$0\leq\alpha<1$. From these figures, one can see that the
behaviour of the diagrams differ as the nonlinear parameter
$\beta$  decreases. This implies that in a very strong nonlinear
regime, the nonlinearity nature of the theory plays a crucial
role. When $\alpha\rightarrow1$, the critical temperatures in
different type of electrodynamic fields toward each others but it
is completely unlike the critical pressure. As one see in Fig.
\ref{fig10}, for $\alpha\rightarrow 1$, the critical pressures
become more different.

\begin{figure}
\centering \subfigure[
$\beta=1$]{\includegraphics[scale=0.4]{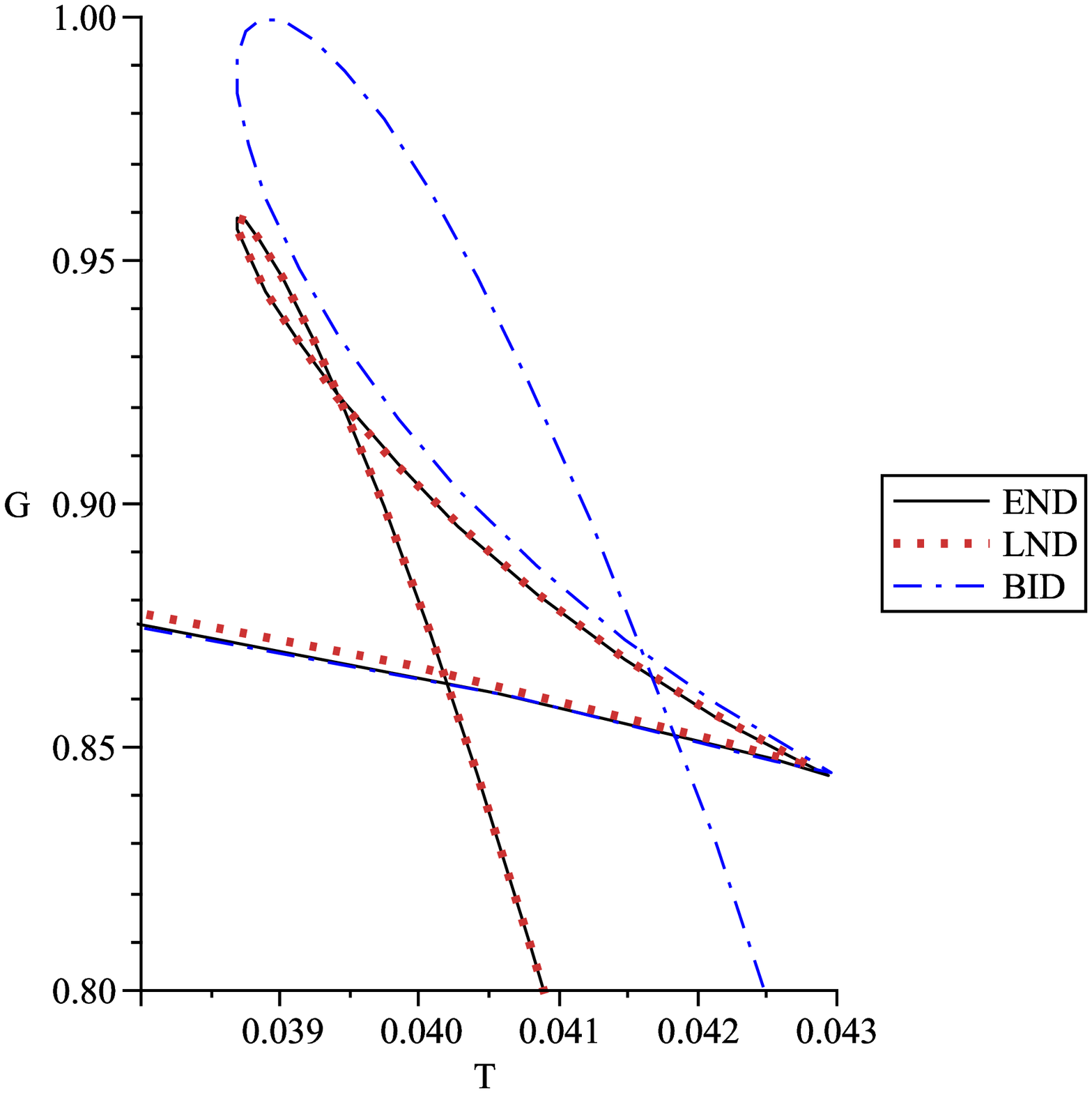}\label{fig5a}}
\hspace*{.1cm} \subfigure[ $\beta=0.25$
]{\includegraphics[scale=0.4]{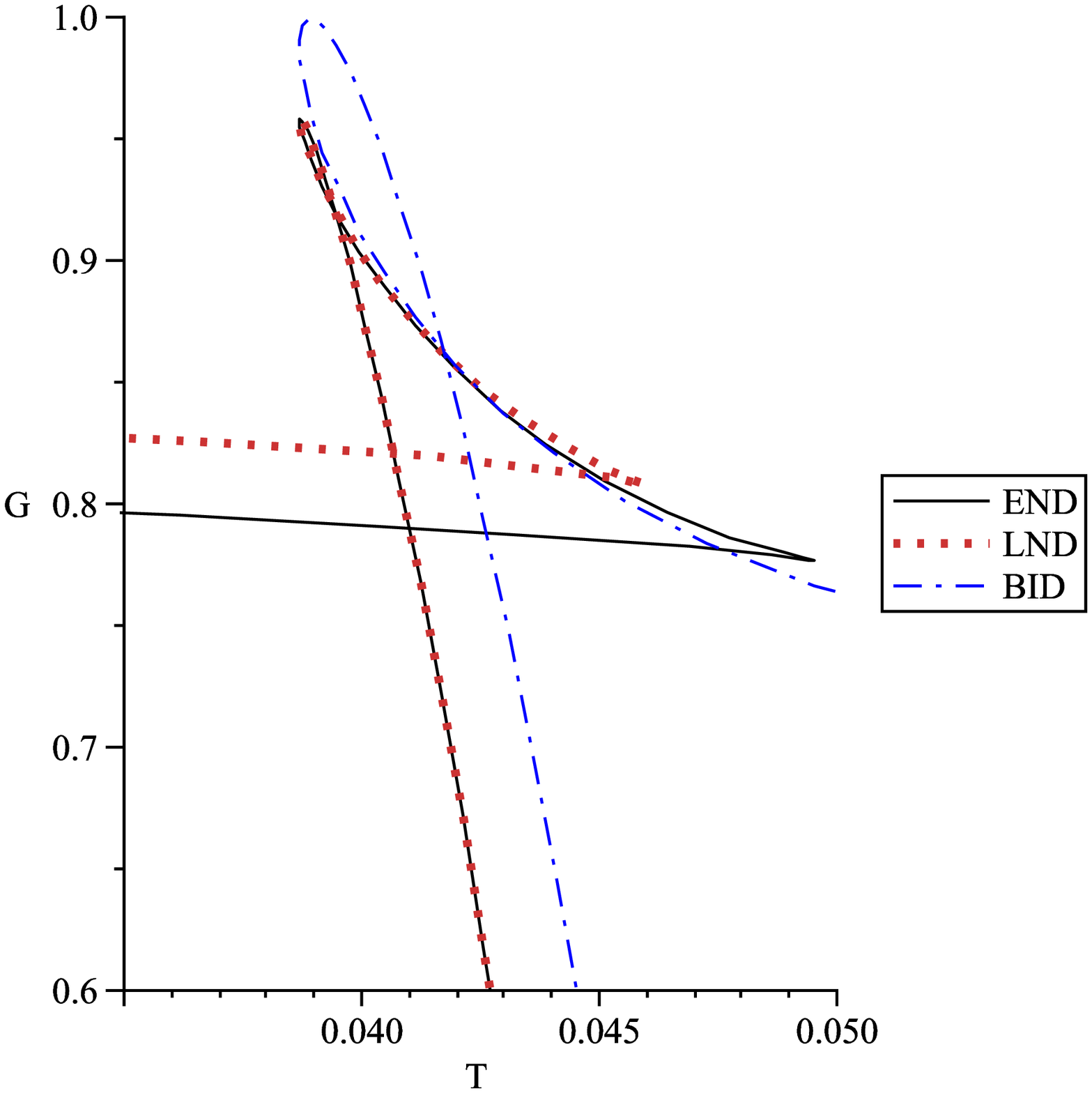}\label{fig5b}}\caption{Gibbs
free energy versus $T$ of dilaton black holes for $q=b=k=1$ and
$\alpha=0.2$ }\label{fig5}
\end{figure}

\begin{figure}
\centering \subfigure[ $\beta=0.5$ and
$T=0.1$]{\includegraphics[scale=0.4]{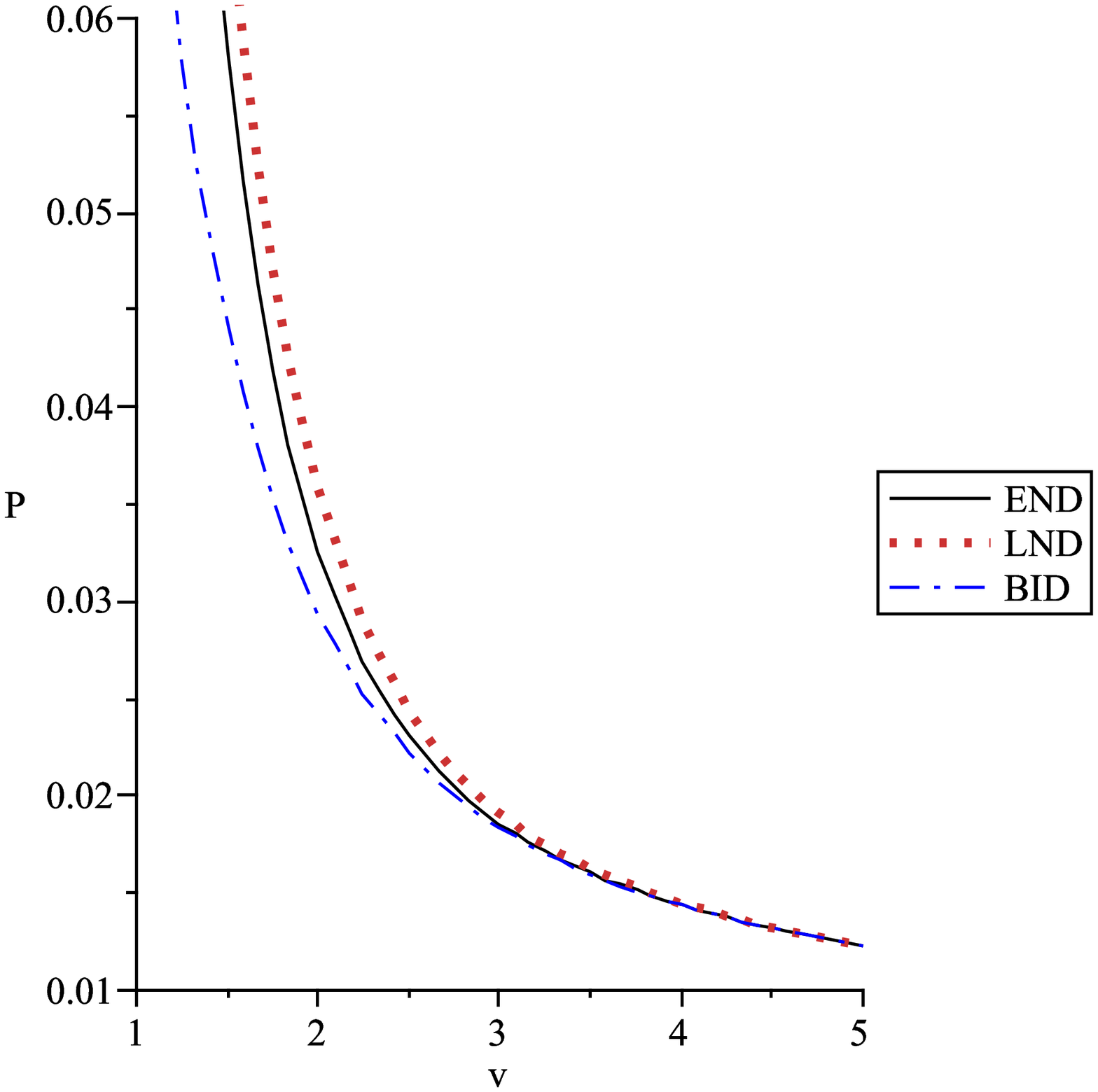}\label{fig6a}}
\hspace*{.1cm} \subfigure[ $\beta=0.25$ and $T=0.3$
]{\includegraphics[scale=0.4]{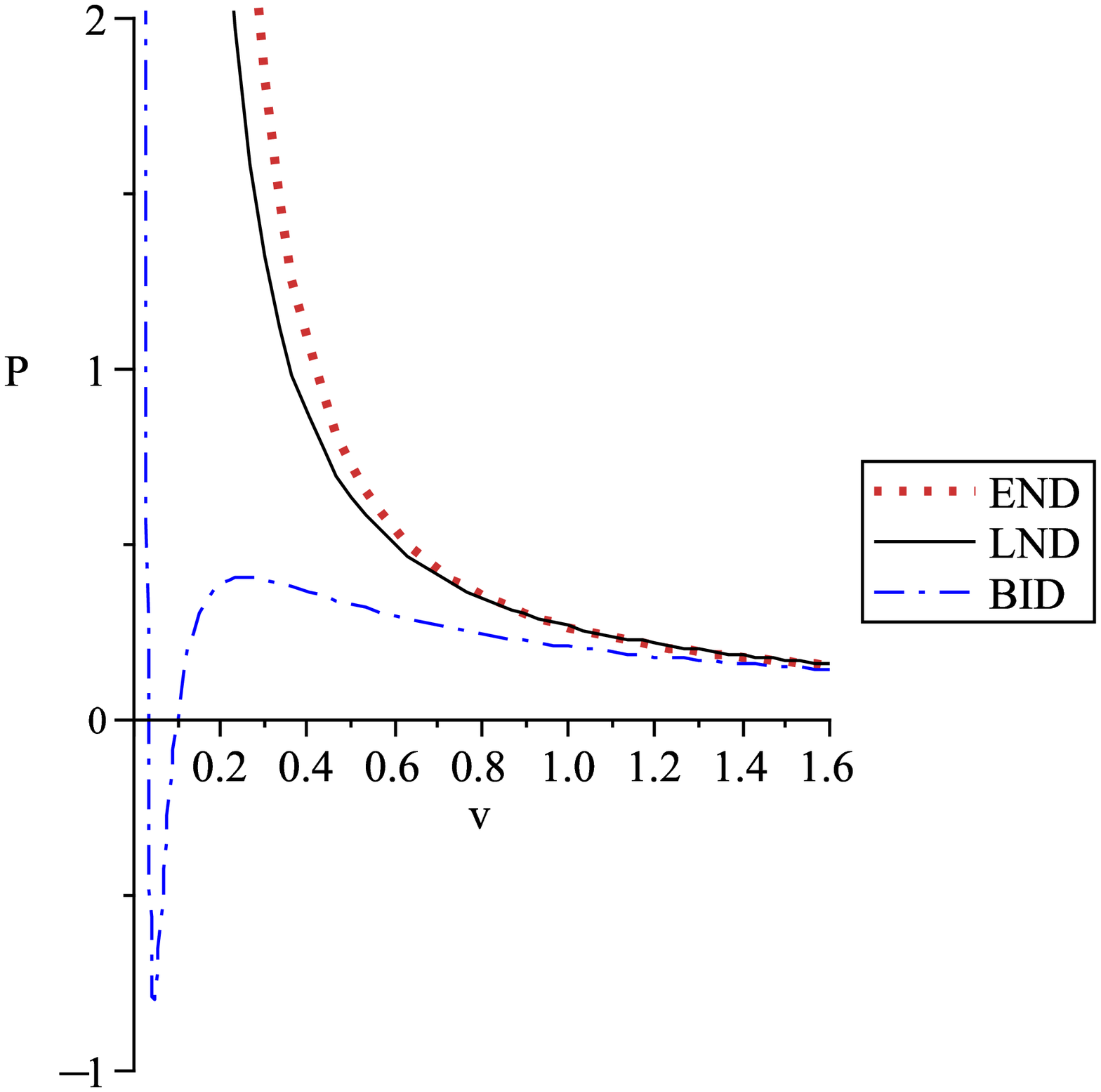}\label{fig6b}}\caption{P-v
diagram of dilaton black holes with different electrodynamics.
Here we have taken $\alpha=0.3$ and $q=b=k=1$.}\label{fig6}
\end{figure}

\begin{figure}
\centering \subfigure[
$\beta=0.7$]{\includegraphics[scale=0.4]{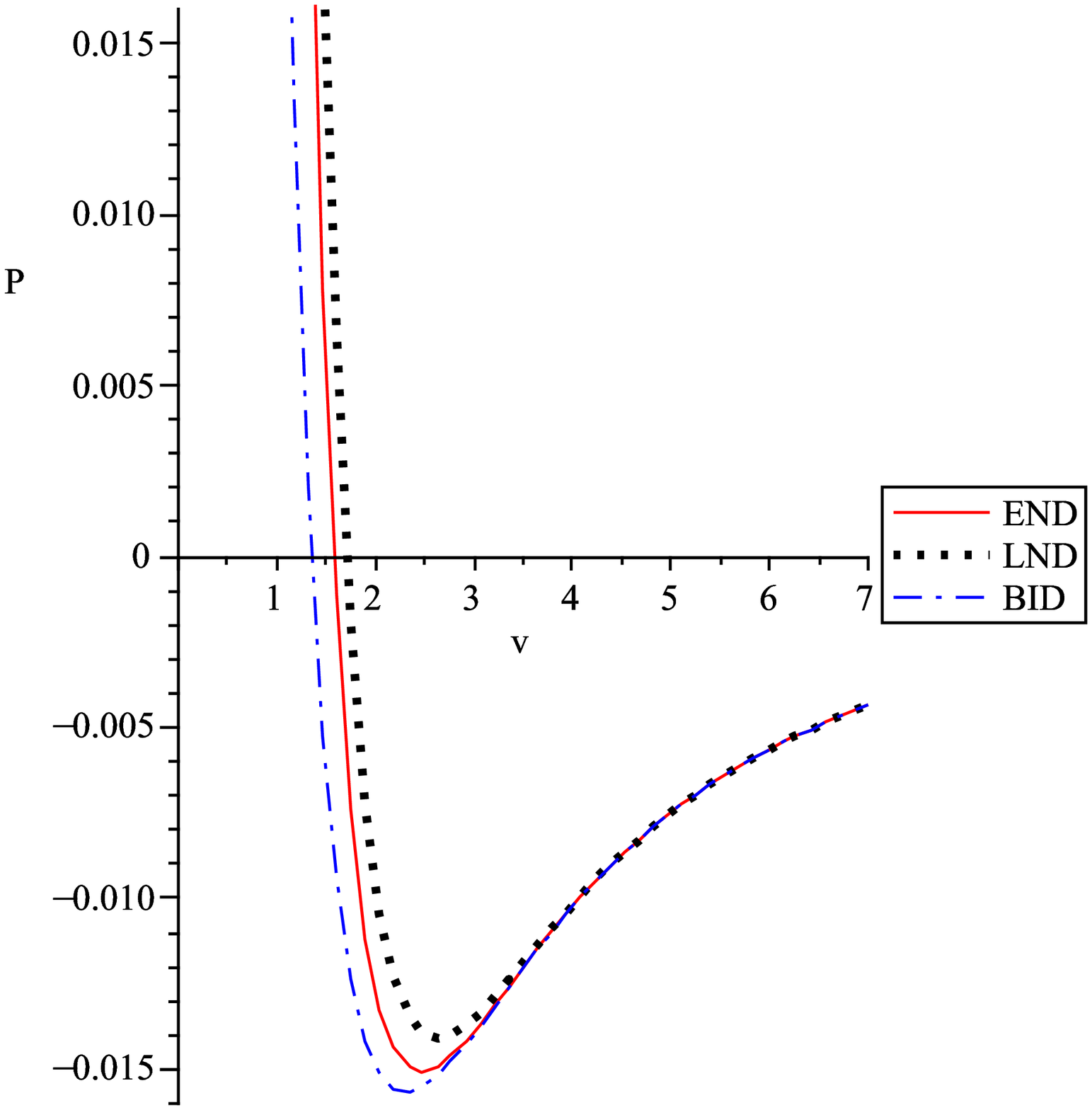}\label{fig7a}}
\hspace*{.1cm} \subfigure[ $\beta=0.4$
]{\includegraphics[scale=0.4]{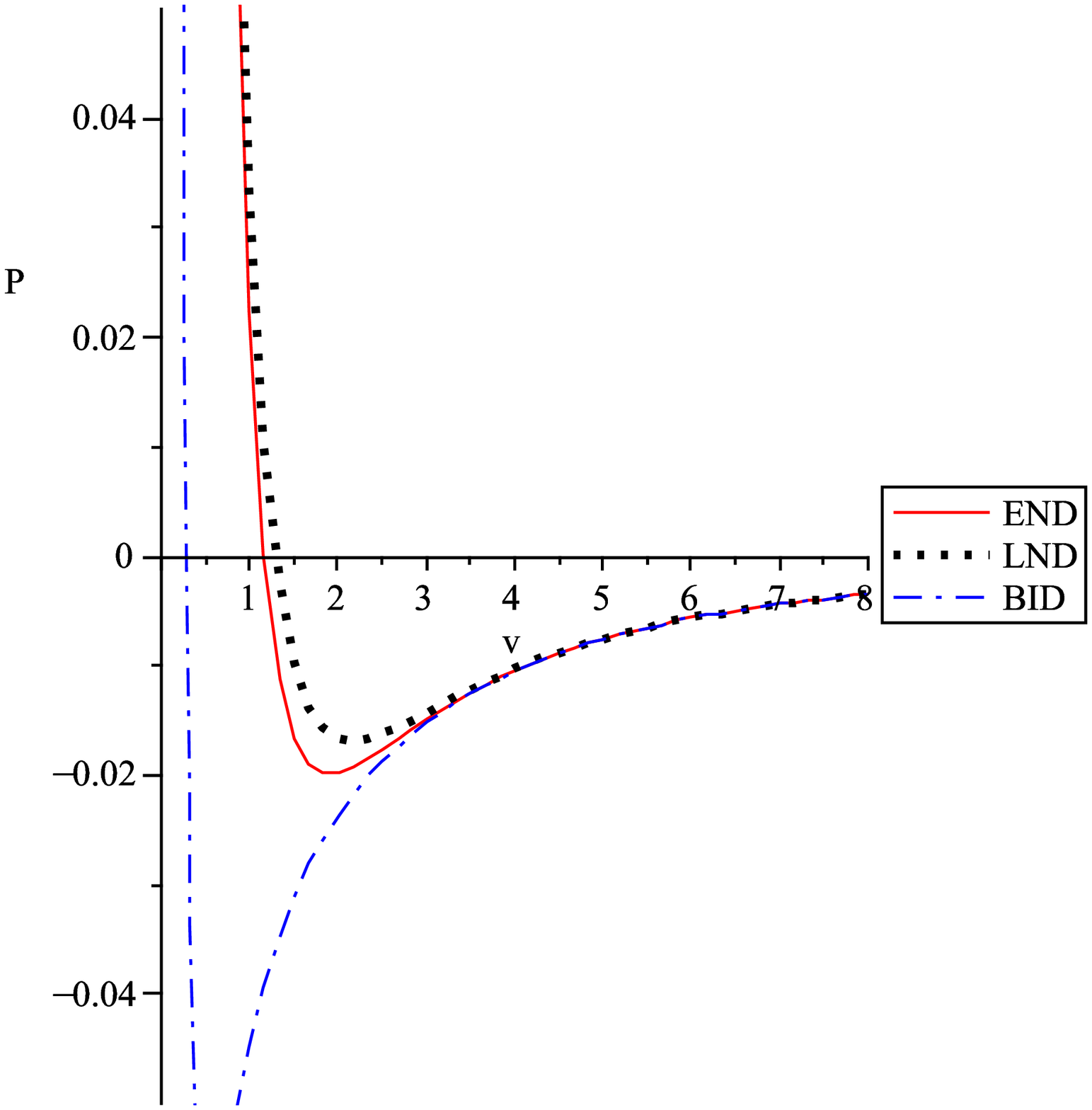}\label{fig7b}}\caption{$P-v$
diagram of dilaton black holes in the same temperature  $T=0.001$
and different $\beta$. Here we have fixed $\alpha=0.3$ and
$q=b=k=1$. }\label{fig7}
\end{figure}

As we already pointed out, although it is hard to calculate the
critical quantities analytically for arbitrary $\beta$, however it
is quite possible to plot the related diagrams for different
$\beta$. We study Gibbs free energy and $P$-$v$ behaviour in Figs.
\ref{fig5} and \ref{fig6}, to see the the difference between the
nonlinear theories we have considered. It is clear from these
diagrams that the behavior of END, LND and BID black holes is very
similar when $T$ or $\beta$ are large enough. As one expects, in
the same $T$, the difference between diagrams increase as $\beta$
decreases (see Fig. \ref{fig7}).

\begin{figure}[H]
\centering
\subfigure[$\alpha=0$]{\includegraphics[scale=0.4]{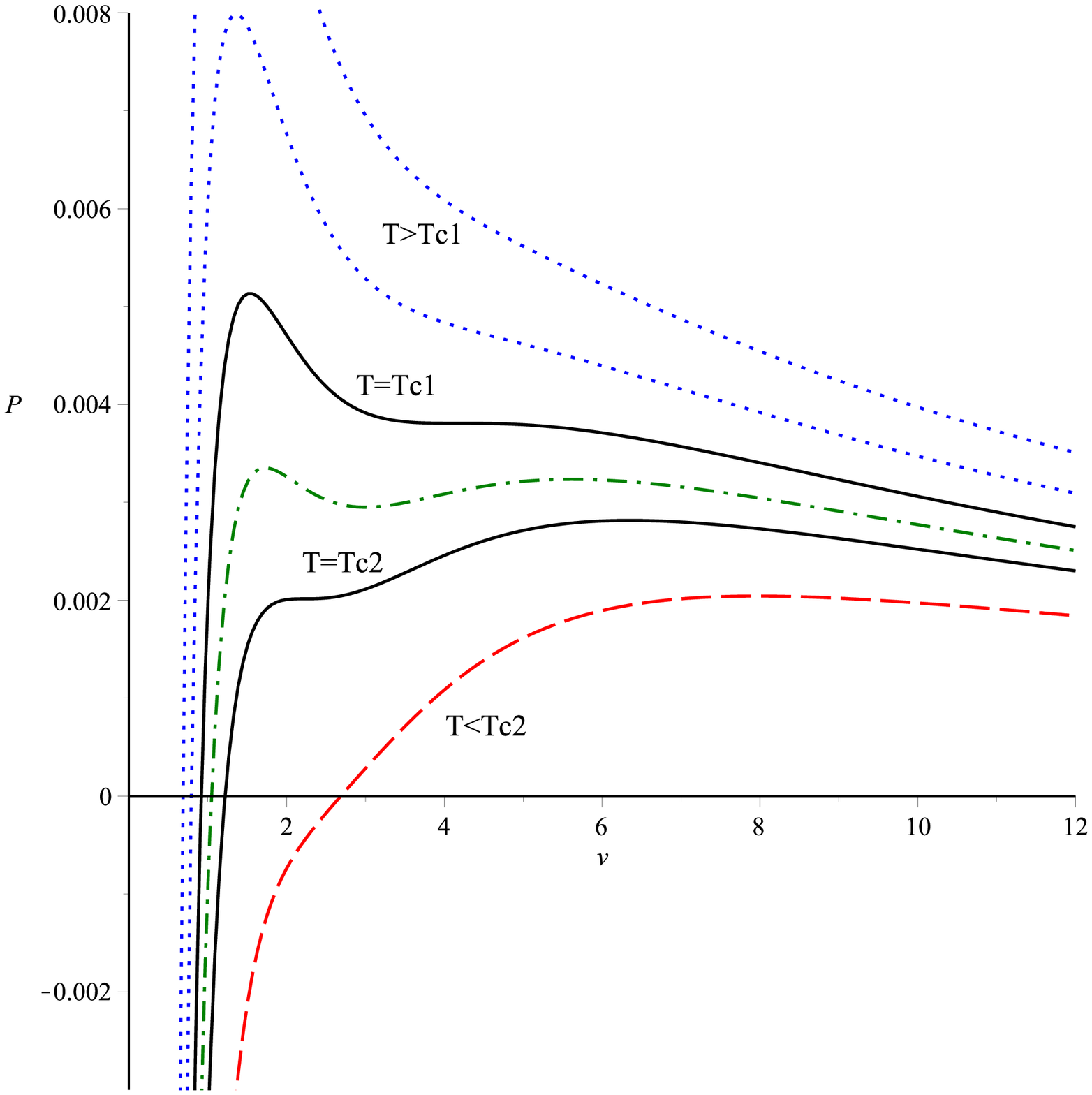}\label{fig11a}}
\hspace*{.1cm} \subfigure[$\alpha=0.1$
]{\includegraphics[scale=0.4]{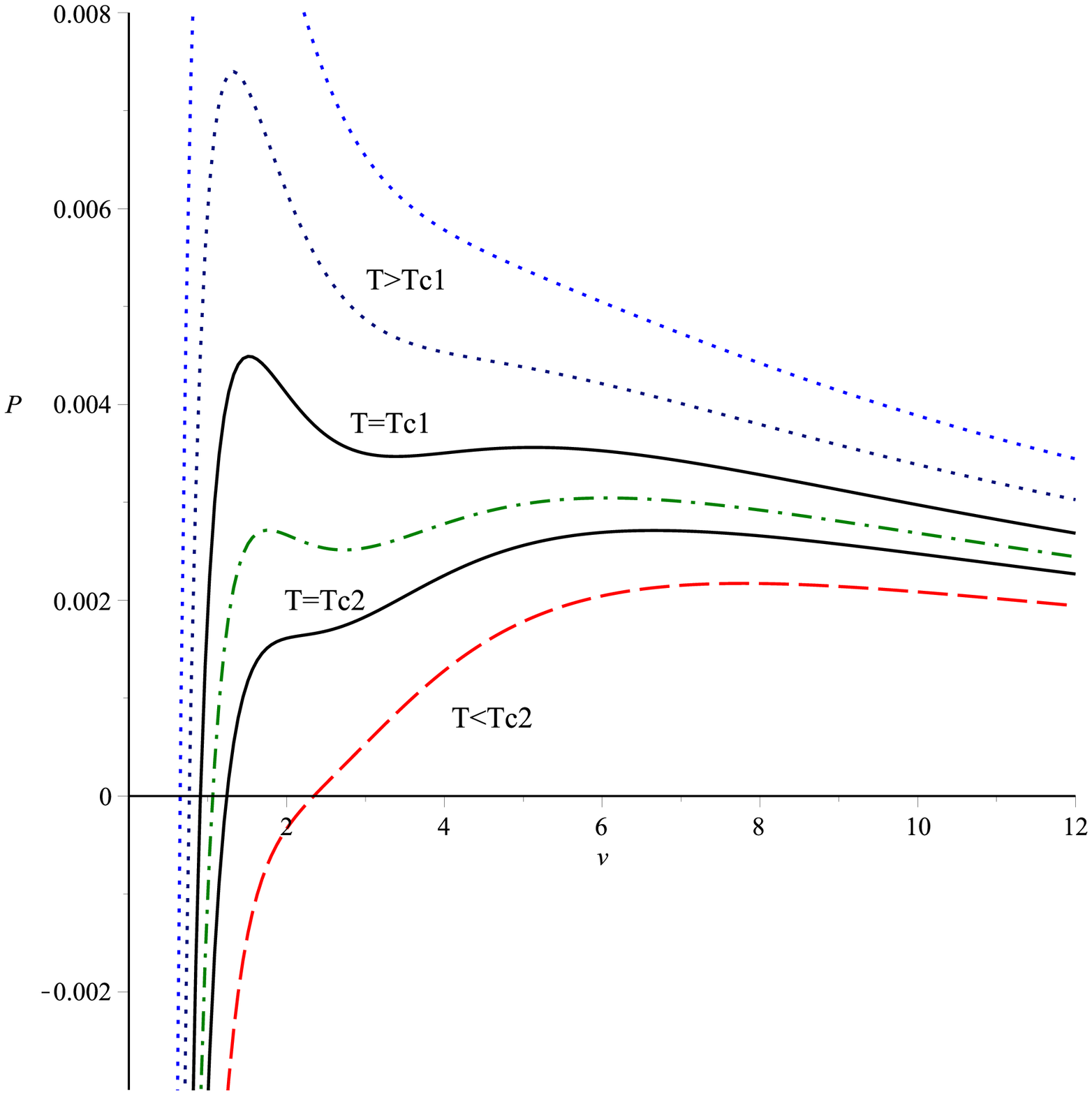}\label{fig11b}}\hspace*{.1cm}
\subfigure[$\alpha=0.25$
]{\includegraphics[scale=0.4]{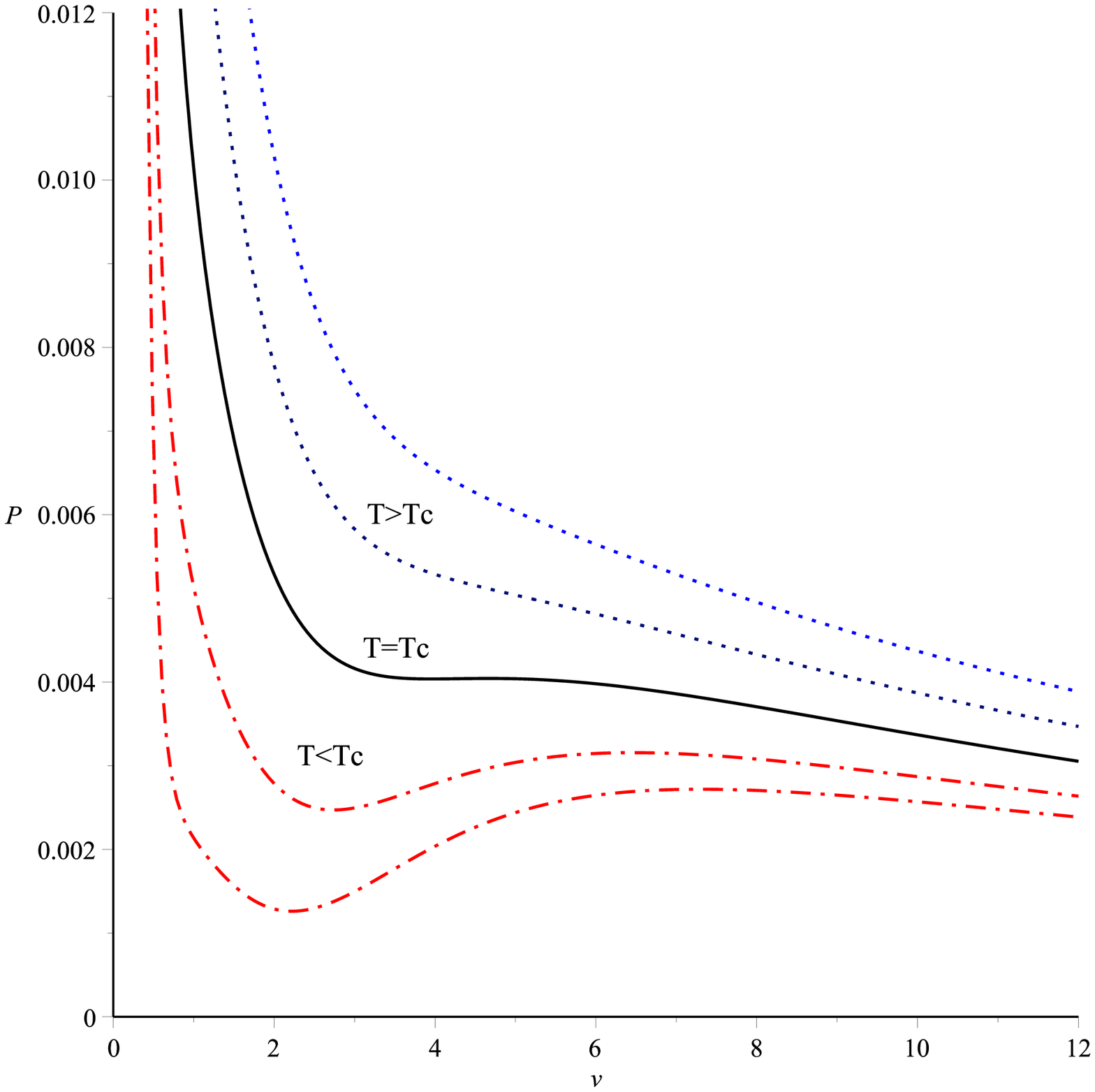}\label{fig11c}}\caption{$P-v$
diagram for BID black holes. Here we have fixed $\beta=0.45$,
$q=b=1$ and  $k=1$.}\label{fig11}
\end{figure}
{It was  extensively argued in \cite{MannBI} that in the absence
of dilaton field, black hole with BI nonlinear electrodynamics may
have two, one or zero critical points which depends on the
strength of nonlinear and charge parameters. For BID black holes,
only for small values of dilaton-electromagnetic coupling $\alpha$
one may see second critical point. Interestingly enough, as
dilaton parameter $\alpha$ increases, the second critical point
disappears. As an example, we compare $P-v$ diagrams of BID black
holes for three values of dilaton coupling $\alpha$ in Fig.
\ref{fig11}. It is clear from these diagrams that in the absence
of dilaton field (Fig. \ref{fig11a}) or for weak dilaton field
(Fig. \ref{fig11b}), there are two critical points but when
dilaton field increases (Fig. \ref{fig11c}) the second critical
point vanish and we have only one critical point. In the other
types of nonlinear electrodynamics such as Logarithmic,
Exponential or Power-law Maxwell fields, the second critical point
is never seen neither in the absence nor in the presence of
dilaton field. Also it is worthwhile to mention that for very
small value of nonlinear parameter $\beta$ there is not any
critical point in all types of above electrodynamics.}
%%%%%%%%%%%%%%%%%%%%%%%%%%%%%%%%%%%%%%%%%%%%%%%%%%%%%%%%
\section{Zeroth order phase transition} \label{PT}
Let us emphasize that the observed phase transition in the
previous sections which were similar to the Van der Walls phase
transition is called the first order phase transition in the
literature. It occurs where Gibbs free energy is continuous, but
its first derivative respect to the temperature and pressure is
discontinuous. Now we want to mention that another interesting
type of phase transition happens in the certain range of the
metric parameters. This discontinuity in Gibbs free energy known
as \textit{zeroth order} phase transition which is observed in
superfluidity and superconductivity \cite{20MrDeh}. It is
important to note that, due to this transition, the response
functions of black holes thermodynamics diverge e.g. isothermal
compressibility. Recently, \textit{zeroth order} phase transition
was observed in the context of Einstein-Maxwell-dilaton black
holes \cite{MrDeh}. It was confirmed that the presence of dilaton
field plays a crucial role for such a phase transition
\cite{MrDeh}. Indeed, there is a direct relation exists between
the zeroth-order portion of the transition curve and dilaton
parameter $\alpha$ \cite{MrDeh}. In other words, we have no zeroth
order phase transition for Einstein-Maxwell (Reissner-Nordstrum)
black holes. Moreover, for nonlinear BI electrodynamics, it was
shown that a zeroth order phase transition may occur even in the
absence of dilaton field \cite{R.Mann}, which means that the
nonlinearity of the gauge field can also cause a zeroth order
phase transition in black holes thermodynamics.

\begin{figure}
    \centering \includegraphics[scale=0.4]{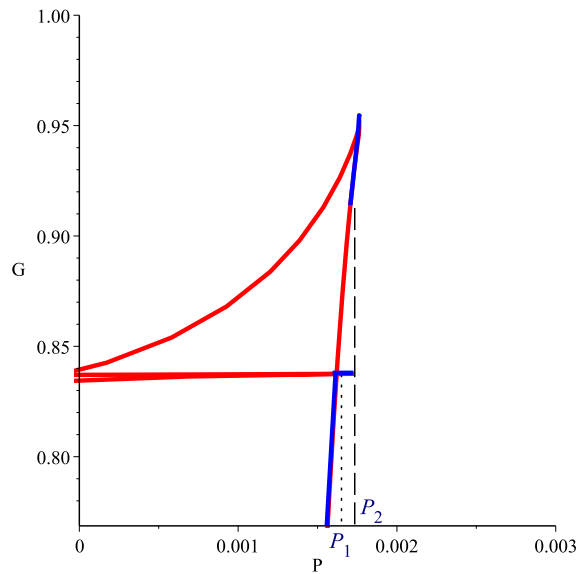}\caption{Gibbs free
        energy for BID black holes versus pressure for $T=0.732 T_c$. Here
        we have fixed $q=b=1$, $\alpha=0.025$, $\beta=0.45$.}\label{fig12}
\end{figure}

\begin{figure}[H]
    \centering \subfigure[BID black
    hole]{\includegraphics[scale=0.4]{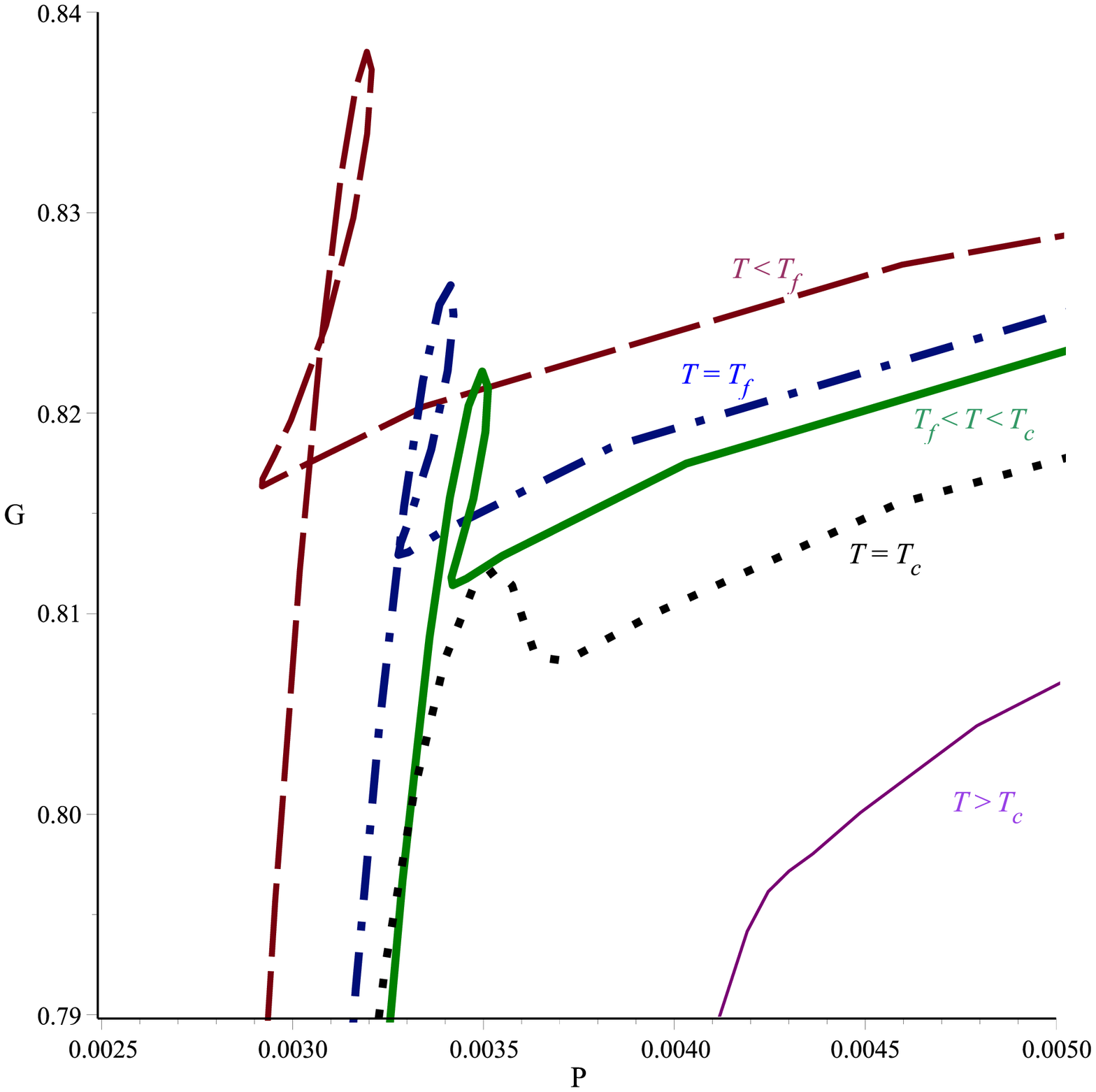}\label{fig13a}}
    \hspace*{.2cm} \subfigure[END black hole
    ]{\includegraphics[scale=0.4]{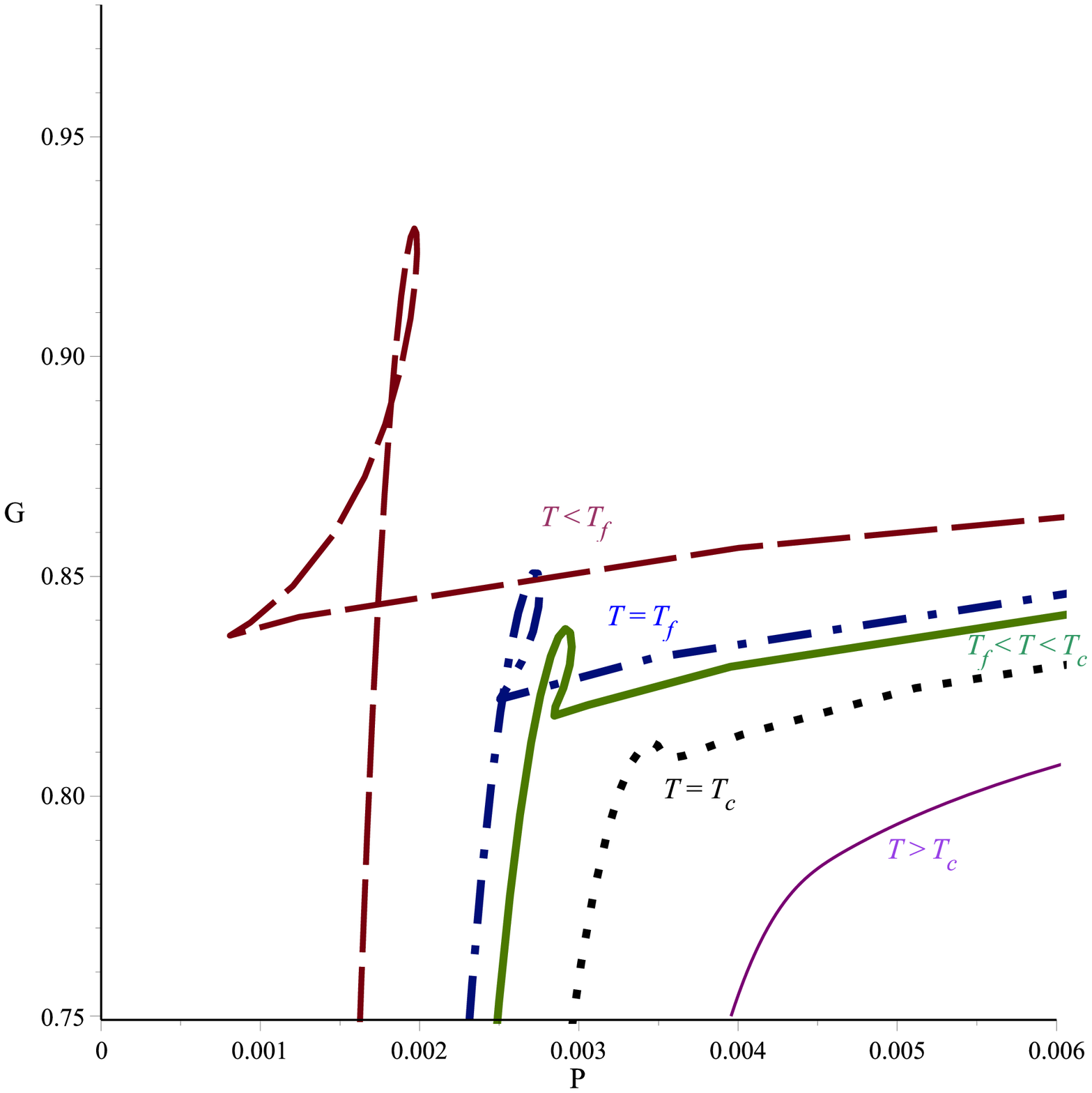}
        \label{fig13b}}\hspace*{.2cm} \subfigure[LND black hole
    ]{\includegraphics[scale=0.4]{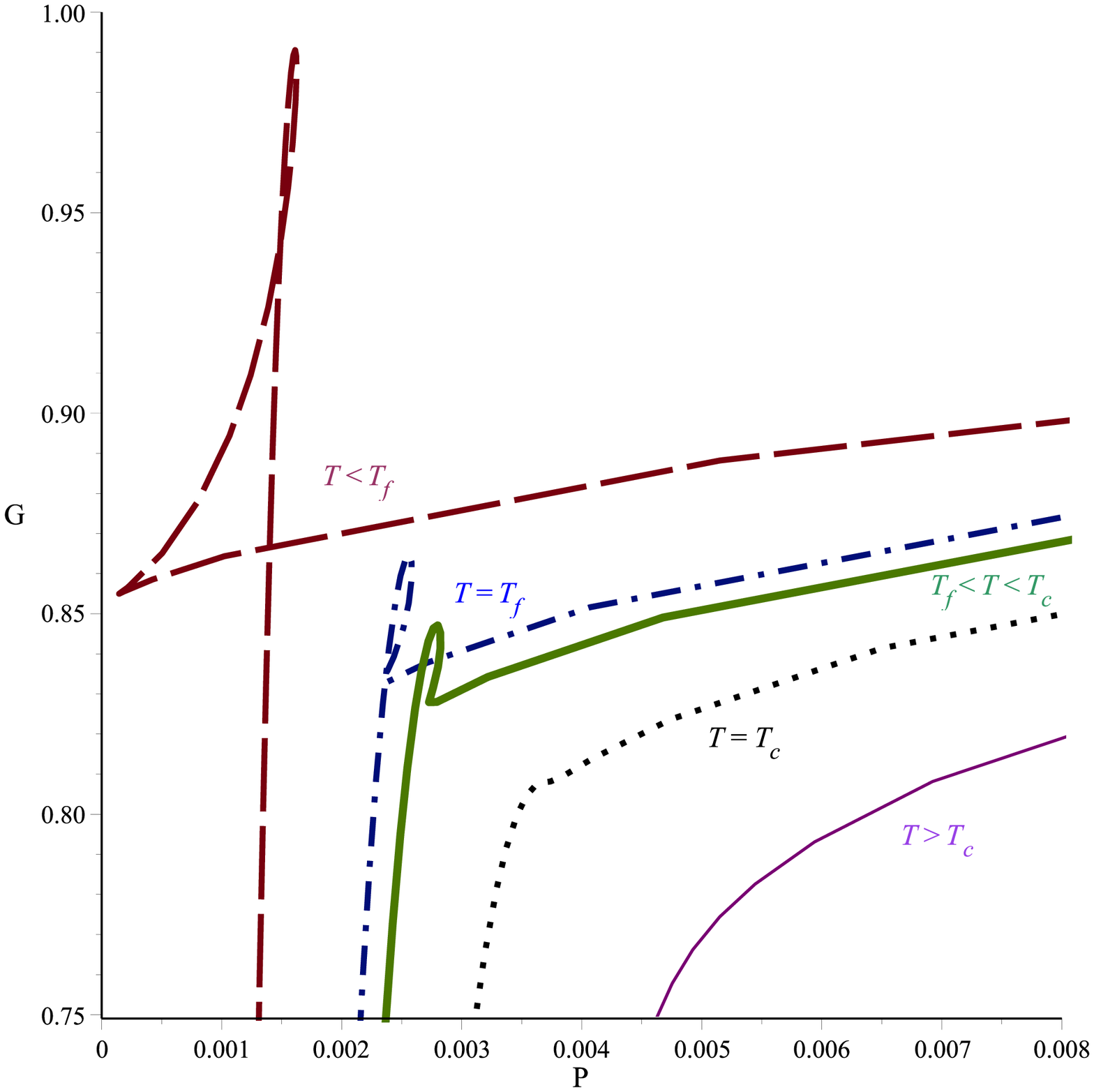}\label{fig13c}}\caption{Gibbs
        free energy versus pressure for $\alpha=0.2$, $\beta=0.45$ and
        $q=k=b=1$.}\label{fig13}
\end{figure}

\begin{figure}[H]
    \centering \subfigure[BID black
    hole]{\includegraphics[scale=0.6]{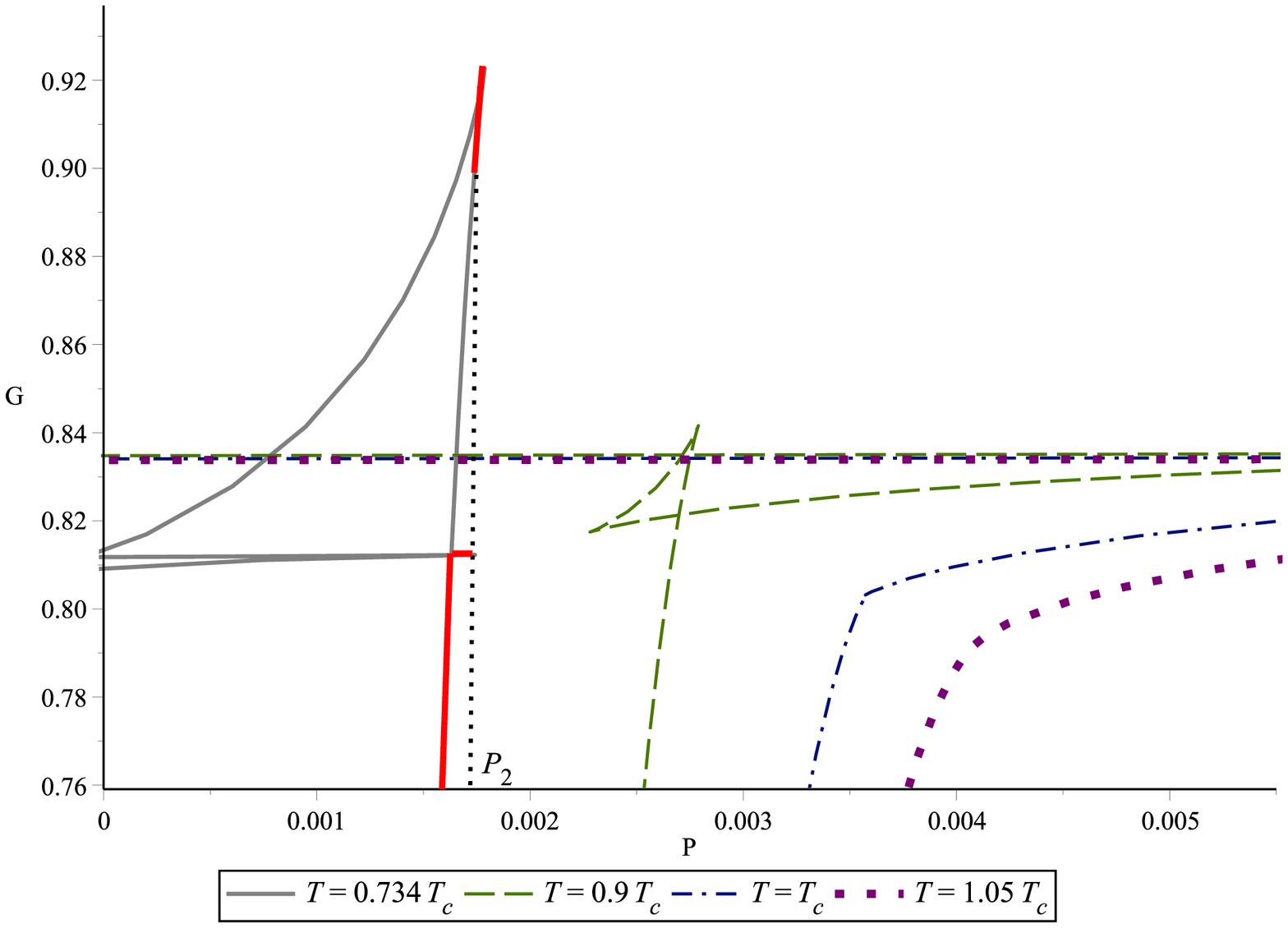}\label{fig14a}}
    \hspace*{.2cm} \subfigure[END black hole
    ]{\includegraphics[scale=0.4]{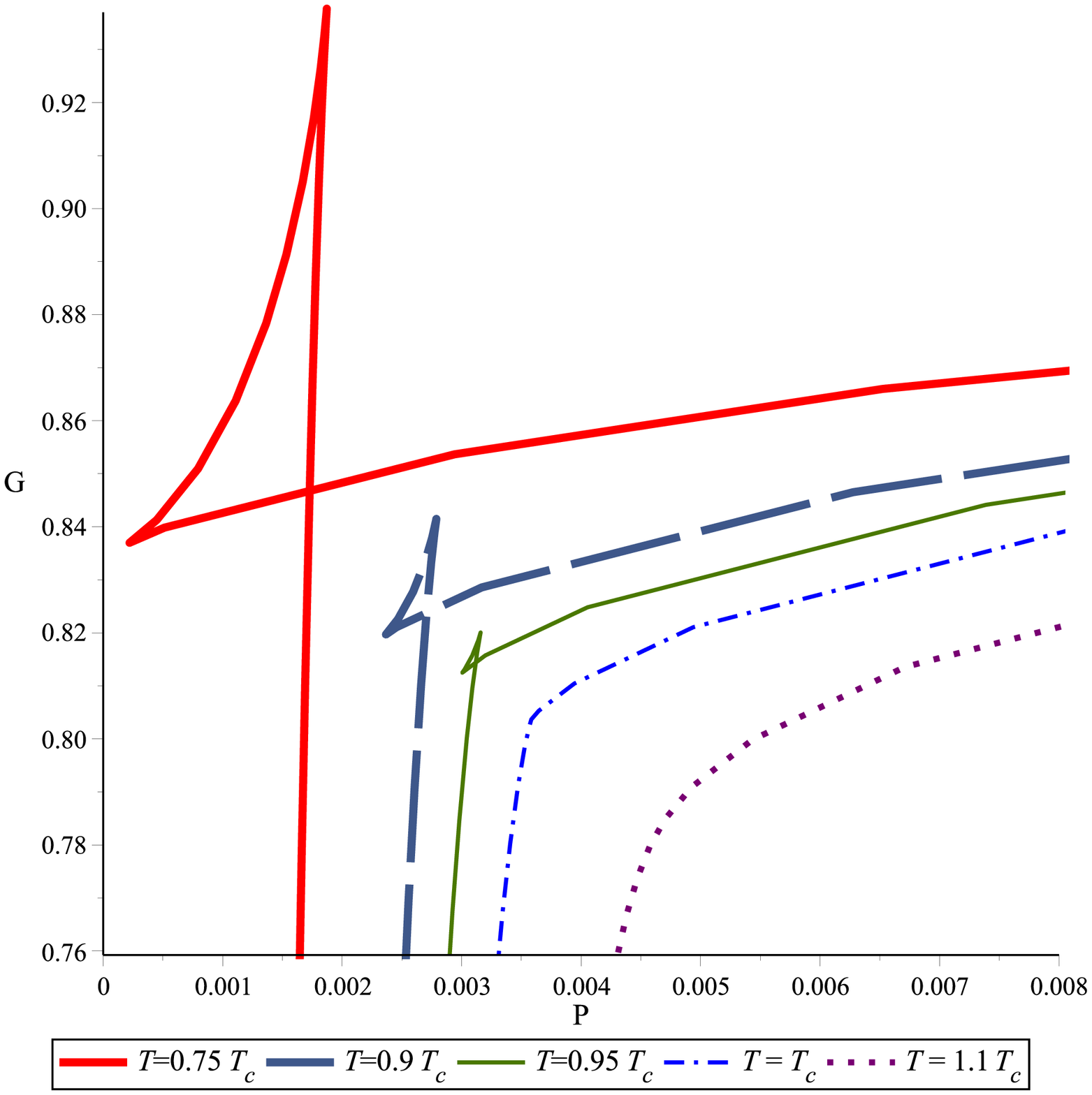}\label{fig14b}}\hspace*{.2cm}
    \subfigure[LND black hole
    ]{\includegraphics[scale=0.4]{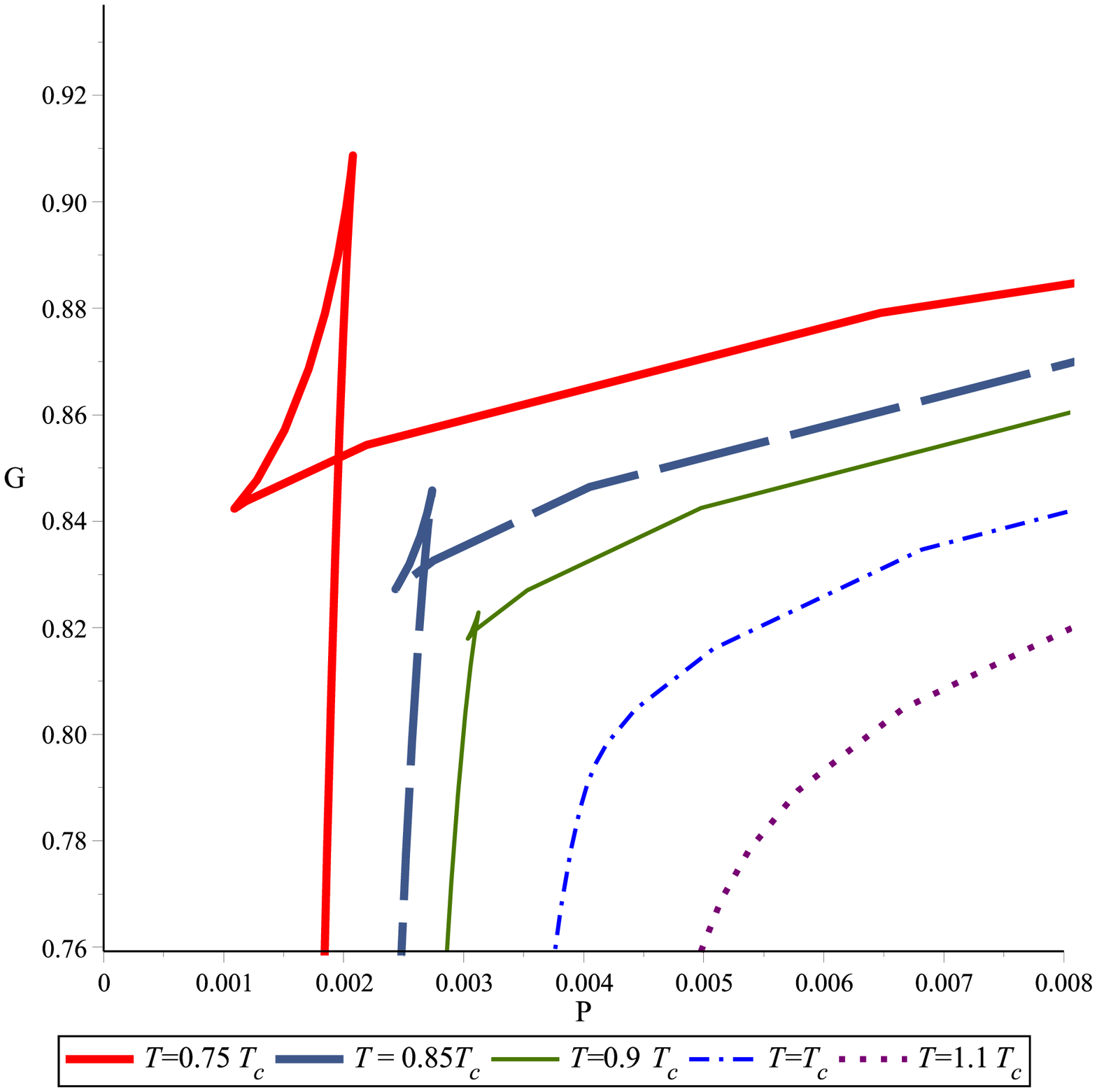}\label{fig14c}}\caption{Gibbs
        free energy versus pressure for $\alpha=0$, $\beta=0.45$ and
        $q=k=b=1$.}\label{fig14}
\end{figure}

Here we would like to explore the possibility to have such a
\textit{zeroth order} phase transition in END and LND black holes,
where both nonlinearity and dilaton field are taken into account.
In order to see the finite jump in Gibbs free energy, we plot the
diagrams of Gibbs free energy respect to the pressure in Figs.
\ref{fig12}, \ref{fig13} and \ref{fig14} for different values of
the metric parameters.

For completeness, we also investigate the phase transition of BID
solutions presented in \cite{Dayyani1}. An interesting case in the
BID theory is plotted in Fig. \ref{fig12}. From this figure, we
see that for a certain values of pressure and especial range of
dilaton field parameter, both zeroth and first order phase
transitions may be observed in one diagram.
% where $P_1$ characterizes the first order phase transition and $P_2$ represents a zeroth order phase transition.
Based on this figure, by increasing the pressure until $P_1$ a
first order transition occurs. For $P>P_1$, Gibbs free energy has
two values and as one can see, the acceptable values of energy are
shown in the blue curve since it includes smaller values of
energy. At point $P_2$, one can see a discontinuity in Gibbs free
energy which demonstrates a zeroth order phase transition.
Also, Fig. \ref{fig13} shows different critical behaviors of dilatonic black
holes in the presence of three nonlinear electrodynamics respect
to the changes in the temperature values when other metric
parameters are fixed. In the case of $T>T_c$, we have no phase
transition. When $T=T_c$, the system experiences a second order
phase transition as we have discussed before. As temperature
decreases to the $T_{\rm f}<T<T_c$ a zeroth order phase
transition is observed. Finally, at $T=T_{\rm f}$ the first order
phase transition occurs. It is worth mentioning that this behavior
is repeated in the Gibbs free energy of all three types of black
holes in the presence of nonlinear electrodynamics and non-zero
values of dilaton field.

It is important to note that by looking at Fig. \ref{fig14}, one
may wonder that, for fixed values of the parameters, and in the
absence of dilaton field ($\alpha=0$), we do not observe zeroth
order phase transition in END and LND theories. This is in
contrast to the BID theory where a zeroth order phase transition
is occurred in the small range of nonlinear parameters $\beta$
even in the absence of dilaton field (see Fig. \ref{fig14a}). In
this figure, the red portion curve shows this behavior as we
explained in close-up Fig. \ref{fig12}. It is one of the main
difference between these three nonlinear electrodynamics, which
implies that their behavior in case of small values of $\beta$
completely differ. This indicate that, while the nonlinearity can
lead to zeroth order phase transition in BI theory, it is not the
case for EN and LN theories. In other words, the presence of the
dilaton field plays a crucial role for occurring zeroth order
phase transition in the context of END and LND electrodynamics.

%%%%%%%%%%%%%%%%%%%%%%%%%%%%%%%%%%%%%%%%%%%%%%%%%%%%%%%%%%%%%%%%%%%%%%%%%%
\section{Closing remarks}\label{Clos}
In this paper, we have studied critical behavior and phase
transition of Exponential and Logarithmic nonlinear
electrodynamics in the presence of dilaton field, which we labeled
them as END and LND, respectively. We extended the phase space by
considering the cosmological constant and nonlinear parameter as
thermodynamic variables. We introduced common conditions to find
solution in both theories, such as potential, metric and etc. We
have investigated these tow nonlinear theories, separately. As the
expansion of END Lagrangian for large nonlinear parameter,
$\beta$, and BID is exactly the same, it is expected that their
critical behavior be the same, in the limit of
$\beta\rightarrow\infty$. We continued our calculation by
obtaining equation of state of END black holes. We observed that
$P-v$ diagrams of this theory are similar to those of Wan der
Waals gas. By applying the approach of Wan der Waals gas to find
out the critical point, we concluded that this point is exactly
the same as in BID black holes. Besides, the Gibbs free energy
diagram confirmed the existence of phase transition and finally
critical exponents were obtained which are exactly the same as the
mean field theory.

We also investigated the critical behaviour of LND black holes.
Again, for $\beta \rightarrow \infty$, the series expansion of LND
Lagrangian is similar to END and BID cases, so one expects that
critical behavior of this theory to be similar to BID and END
theories in this limit. Our calculations confirmed that the
critical behavior of LND theory is exactly the same as those of a
Wan der Waals gas system.

{It is important to note that although the critical behaviour of
END and LND electrodynamics, in the limit of large nonlinear
parameter $\beta$, is similar to BID black holes explored in Ref.
\cite{Dayyani1}, however, for small value of $\beta$, the
situation quite differs and the behaviour of these three type of
nonlinear electrodynamics are completely different. For example,
it was argued in \cite{MannBI} that BI black holes may have two,
one or zero critical points, however, this behaviour is not seen
for Logarithmic and Exponential, namely the second critical point
is never seen in the absence/presence of dilaton field.}

{We also investigated the phase transition of END and LND black
holes. In addition to the usual critical (second-order) as well as
the first-order phase transitions in END and LND black holes, we
observed that a finite jump in Gibbs free energy is generated by
dilaton-electromagnetic coupling constant, $\alpha$, for a certain
range of pressure. This novel behavior indicates a small/large
black hole \emph{zeroth-order} phase transition in which the
response functions of black holes thermodynamics diverge. It is
worthy to note that for temperature in the range $T_{\rm
f}<T<T_c$, a discontinuity occurs in the Gibbs free energy diagram
which leads to zeroth order phase transition. We find out that in
the absence of dilaton field, we do not observe zeroth order phase
transition in END and LND theories. This is in contrast to the BI
theory where a zeroth order phase transition is occurred in the
small range of nonlinear parameters $\beta$ even in the absence of
dilaton field. We conclude that, while in BI black holes, the
nonlinearity can lead to zeroth order phase transition, it is not
the case for EN and LN black holes. In other words, the presence
of dilaton field plays a crucial role for occurring zeroth order
phase transition in the context of EN and LN electrodynamics.}

{Finally, we would like to mention that the jump in the Gibbs free
energy is observed for three types of dilatonic nonlinear
electrodynamics, namely BID, END and LND. However, in the absence
of dilaton field, a zeroth order phase transition occurs only for
BI black holes, which means that the nonlinearity is responsible
for this phase transition. However, for LND and END black holes,
it seems the dilaton field is responsible for this type of zeroth
order phase transition. Albeit, for BID theory, both dilaton field
as well as nonlinear electrodynamics can lead to zeroth order
phase transition. This behaviour and the physical reasons behind
it, need further investigations in the future studies.}
%%%%%%%%%%%%%%%%%%%%%%%%%%%%%%%%%%%%%%%%%%%%%%%%%%%%%%%%%%%%%%%%%
\acknowledgments{We are grateful to the referee for constructive
comments which helped us improve our paper significantly. We also
thank Shiraz University Research Council. The work of AS been
supported financially by Research Institute for Astronomy and
Astrophysics of Maragha, Iran.}
%%%%%%%%%%%%%%%%%%%%%%%%%%%%%%%%%%%%%%%%%%%%%%%%%%%%%%%%%%%%%%%%%%%


\begin{thebibliography}{99}
\bibitem{Bek} J.D. Bekenstein, Phys. Rev. D \textbf{7}, 2333 (1973); J.D.
Bekenstein, Phys. Rev. D \textbf{9}, 3292 (1974).

\bibitem{Haw} S.W. Hawking, Commun. Math. Phys. \textbf{43}, 199 (1975);
S.W. Hawking, Phys. Rev. D \textbf{13}, 191 (1976).

\bibitem{hawking-page} S. Hawking and D. N. Page, Commun. Math. Phys. {\bf87},  577
(1983). %Thermodynamics of Black Holes in anti-De Sitter Space%

\bibitem{Do1} B. P. Dolan, Class. Quant. Grav. \textbf{28}, 235017 (2011).
%\textit{Pressure and volume in the first law of black hole thermodynamics}%

\bibitem{Ka} D. Kastor, S. Ray, and J. Traschen, Class. Quant. Grav. \textbf{%
26}, 195011 (2009).
%\textit{Enthalpy and the Mechanics of AdS Black Holes}%

\bibitem{Do2} B. Dolan, Class. Quant. Grav. \textbf{28}, 125020 (2011).
%\textit{The cosmological constant and black hole equation of state},%

\bibitem{Do3} B. P. Dolan, Phys. Rev. D \textbf{84}, 127503 (2011).
%\textit{Compressibility of rotating black holes},%

\bibitem{Ce1} M. Cvetic, G. W. Gibbons, D. Kubiznak, and C. N. Pope, Phys.
Rev. D \textbf{84}, 024037 (2011).
%\textit{%Black hole enthalpy and an entropy inequality for the thermodynamic volume},%

\bibitem{Ur} M. Urana, A. Tomimatsu, and H. Saida, Class. Quant. Grav.
\textbf{26}, 105010 (2009).
%\textit{Mechanical first law of black hole spacetime with cosmological constant and its application to Schawarzchils-de Sitter spacetime}%

\bibitem{MannRN} D. Kubiznak and R. B. Maan, J. High Energy Physics, \textbf{%
07}, 033 (2012). %\textit{P-V criticality of charged Ads black holes }%

\bibitem{MannBI} Sh. Gunasekaran, D. Kubiznak and R. B. Mann, JHEP, \textbf{11}, 110 (2012).
%\textit{%Extended phase space thermodynamics for charged and rotating black holes and Born-Infeld vacuum polarization}%


\bibitem{HV} S. H. Hendi, M. H. Vahidinia, Phys. Rev. D {\bf88}, 084045
(2013). %Extended phase space thermodynamics and P-V criticality of black holes with nonlinear source%

\bibitem{Hendi1} S. H. Hendi, S. Panahiyan, B. Eslam Panah, Int. J. Mod. Phys. D Vol. {\bf25}, No. 1, 1650010
(2016). %P-V  criticality and geometrothermodynamics of black holes with Born-Infeld type nonlinear electrodynamics%

\bibitem{GB1} S.-W. Wei and Y.-X. Liu, Phys. Rev. D {\bf87}, 044014
(2013). %Critical phenomena and thermodynamic geometry of charged Gauss- Bonnet AdS black holes%

\bibitem{GB2} De. Zou, Y.i Liu, B. Wang, Phys.Rev. D {\bf90}, 044063
(2014). %Critical behavior of charged Gauss-Bonnet AdS black holes in the grand canonical ensemble%


\bibitem{Lovelock} A. Frassino, D. Kubiznak, R. Mann, and F. Simovic, JHEP {\bf09}, 080
(2014);\\ J. X. Mo, W. B. Liu, Eur. Phys. J. C. \textbf{74}, 2836
(2014). %\textit{P-V Criticality of Topological Black Holes in Lovelock-Born-Infeld Gravity}%

\bibitem{Sherkat} M. B. Jahani Poshteh, B. Mirza and Z. Sherkatghanad, Phys.
Rev. D 88, 024005 (2013).
%\textit{Phase transition, critical behavior, and critical exponents of Myers-Perry black holes}%

\bibitem{Sherkat1} Z. Sherkatghanad, B. Mirza, Z. Mirzaeyan and S. A.
Hosseini Mansoori, arXiv:1412.5028.
%\textit{Critical behaviors and phase transitions of black holes in higher order gravities and extended phase spaces}%

\bibitem{Rabin} R. Banerjee and D. R Roychowdhury, Phys. Rev. D \textbf{85},
044040 (2012);\newline R. Banerjee, D. Roychowdhury, Phys. Rev. D
\textbf{85}, 104043 (2012).
%\textit{Critical behavior of Born Infeld AdS black holes in higher dimensions}%

\bibitem{Zou} De. Ch. Zou, Sh.-J. Zhang and B. Wang, Phys. Rev. D \textbf{89}%
, 044002 (2014).
%\textit{\ Critical behavior of Born-Infeld AdS black holes in the extended phasespace thermodynamics}%

\bibitem{John} C. V. Johnson, Class. Quant. Grav. \textbf{31}, 225005
(2014);\\ C. O. Lee, Phys. Let. B \textbf{09} (2014) 046.
%\textit{The extended thermodynamic phase structure of Taub-Nut and Taub-Bolt}%

\bibitem{Born} M. Born and L. Infeld, %{Foundation of new field theory}%,
Proc. R. Soc. A \textbf{144}, 425 (1934).
\bibitem{Soleng} H. H. Soleng, Phys. Rev. D {\bf52}, 6178 (1995).
\bibitem{Hassaine} M. Hassaine and C. Martinez, Phys. Rev. D {\bf75}, 027502 (2007).
\bibitem{Hendi3} S. H. Hendi, Phys. Lett. B {\bf677}, 123 (2009).

\bibitem{Green} M. B. Green, J. H. Schwarz, and E. Witten, \textit{Superstring
Theory} (Cambridge University Press, Cambridge, England, 1987).
\bibitem{d1} G. W. Gibbons and K. Maeda, %Black Holes and Membranes in Higher Dimensional Theories with Dilaton Fields%, Nucl. Phys. B
298(1988)741
\bibitem{d2} D. Garfinkle, G. T. Horowitz and A. Strominger, %Charged black holes in string theory%,
 Phys. Rev. D {\bf43} (1991) 3140.
\bibitem{CHM} K. C. K. Chan, J. H. Horne and R. B. Mann, %Charged dilaton black holes with unusual asymptotics,%
 Nucl. Phys. B {\bf447} (1995) 441.
\bibitem{d4} G. Clement and C. Leygnac,  Phys. Rev. D {\bf70}  (2004)
084018. %Non-asymptotically flat, non-AdS dilaton black holes,%
\bibitem{d5} C. J. Gao and H. N. Zhang, Phys. Lett. B {\bf612}
(2006) 127. %Topological Black Holes in Dilaton Gravity Theory%

\bibitem{Cai3} R. G. Cai and K. S. Soh, Phys. Rev. D \textbf{59}, 044013
(1999).
\bibitem{neda} M. H Dehghani and N. Farhangkhah,  Phys. Rev. D {\bf 71}, 044008 (2005);\\
 M. H. Dehghani, S. H. Hendi, A. Sheykhi and H. Rastegar Sedehi, JCAP {\bf02} (2007)
020. %Thermodynamics of Rotating Black Branes in Einstein–Born-Infeld-dilaton Gravity,%
\bibitem{Shey3} A. Sheykhi, Phys. Rev. D \textbf{76}, 124025
(2007). %Thermodynamics of charged topological dilaton black holes%


\bibitem{d7} A. Sheykhi, N. Riazi, M. H. Mahzoon,  Phys. Rev. D {\bf74}, 044025
(2006);\\ %Asymptotically nonflat Einstein-Born-Infeld-dilaton black holes with Liouville-type potential%
A. Sheykhi, Phys. Lett. B {\bf662}, 7 (2008);\\ %Topological Born-Infeld-dilaton black holes%
A. Sheykhi, N. Riazi, Phys. Rev. D {\bf75}, 024021 (2007);\\ %Thermodynamics of black holes in (n+1)-dimensional Einstein-Born-Infeld dilaton gravity%
A. Sheykhi, Int. J. Mod. Phys. D {\bf18} 25 (2009). %Thermodynamical properties of topological Born-Infeld-dilaton black holes%

\bibitem{Kamrani} M. H. Dehghani, S. Kamrani, A. Sheykhi, Phys. Rev. D.
\textbf{90}, 104020 (2014). %\textit{P-V criticality of charged dilatonic black holes}%

\bibitem{Dayyani1} M. H. Dehghani, A. Sheykhi and Z. Dayyani, Phys. Rev. D
\textbf{93}, 024022 (2016).
%\textit{%Critical behavior of Born-Infeld dilaton black holes}%

\bibitem{Dayyani2} Z. Dayyani, A. Sheykhi and M. H. Dehghani, Phys  Rev  D \textbf{95}, 084004 (2017), [arXiv:1611.00590].
 %Counterterm method in dilaton gravity and the critical behavior of dilaton black holes with power-Maxwell field%


\bibitem{somayeh} A. Sheykhi and S. Hajkhalili, Phys. Rev. D \textbf{89}, 104019 (2014).
\bibitem{sara} A. Sheykhi, F.Naeimipour, S. M. Zebarjad, Phys. Rev. D \textbf{91}, 124057 (2015).

\bibitem{GW} G. W. Gibbons,  Rev. Mex. Fis. \textbf{49S1} 19, (2003),
[hep-th/0106059]. %\textit{ Aspects of Born-Infeld theory and string/M theory,}%

\bibitem{Lambert} M. Abramowitz and I. A. Stegun, \emph{Handbook of Mathematical
Functions}, Dover, New York, (1972);\\
R. M. Corless, etal., Adv. Computational Math. {\bf5}, 329 (1996).

\bibitem {abot} L. F. Abbott and S. Deser, Nucl. Phys. {\bf B195}, 76 (1982).

\bibitem{enthalpy} D. Kastor, S. Ray, and J. Traschen, Class. Quant. Grav. \textbf{26}, 195011 (2009).

\bibitem{20MrDeh} V. P. Maslov, Math Notes \textbf{76}, 697 (2004).
\bibitem{MrDeh} A. Dehyadegari, A. Sheykhi, A. Montakhab,
[arXiv:1707.05307]. %Novel phase transition in charged dilaton black holes%

\bibitem{R.Mann} S. Gunasekaran, R. B. Mann and D. Kubiznak, JHEP
11, \textbf{110} (2012).
\end{thebibliography}
\end{document}